\documentclass[sigconf,natbib=true]{acmart}
\usepackage{amsmath}
\usepackage{algorithm}
\usepackage{algorithmic}
\usepackage{multirow}
\usepackage{xspace}
\usepackage{subfigure}
\usepackage{enumitem}
\usepackage{makecell}

\newcommand{\method}{CURec\xspace}

\AtBeginDocument{%
  }

\copyrightyear{2026}
\acmYear{2026}
\setcopyright{cc}
\setcctype{by}
\acmConference[CIKM '26]{Proceedings of the 35th ACM International Conference on Information and Knowledge Management}{November 07--11, 2026}{Rome, Italy}
\acmBooktitle{Proceedings of the 35th ACM International Conference on Information and Knowledge Management (CIKM '26), November 07--11, 2026, Rome, Italy}
\acmDOI{10.1145/3799682.3840740}
\acmISBN{979-8-4007-2539-5/2026/11}

\begin{document}

\title[Towards Comprehensible Recommendation with Large Language Model Fine-tuning]{Towards Comprehensible Recommendation\\with Large Language Model Fine-tuning}

\author{Yunze Luo}
\affiliation{
    \institution{School of CS, Peking University}
    \city{Beijing}
    \country{China}}
\email{lyztangent@pku.edu.cn}

\author{Gaode Chen}
\affiliation{
    \institution{Kuaishou Technology}
    \city{Beijing}
    \country{China}}
\email{chengaode19@gmail.com}

\author{Yinjie Jiang}
\affiliation{
    \institution{Kuaishou Technology}
    \city{Beijing}
    \country{China}}
\email{jiangyinjie@kuaishou.com}

\author{Xinghua Zhang}
\affiliation{
    \institution{Unaffiliated}
    \city{Beijing}
    \country{China}}
\email{zxh.zhangxinghua@gmail.com}

\author{Kaigui Bian}
\authornote{Corresponding author.}
\affiliation{
    \institution{School of CS, Peking University}
    \city{Beijing}
    \country{China}}
\email{bkg@pku.edu.cn}

\thanks{Yunze Luo and Kaigui Bian are affiliated with Beijing Key Laboratory of Software and Hardware Cooperative Artificial Intelligence Systems, School of CS, Peking University.}

\renewcommand{\shortauthors}{Yunze Luo, Gaode Chen, Yinjie Jiang, Xinghua Zhang, and Kaigui Bian}
\begin{abstract}
Recommender systems have become increasingly ubiquitous in daily life. While traditional recommendation approaches primarily rely on ID-based representations or item-side content features, they often fall short in capturing the underlying semantics aligned with user preferences (\textit{e.g.}, recommendation reasons for items), leading to a semantic-collaborative gap. 
Recently emerged LLM-based feature extraction approaches also face a key challenge: how to ensure that LLMs possess recommendation-aligned reasoning capabilities and can generate accurate, personalized reasons to mitigate the semantic-collaborative gap.
To address these issues, we propose a novel Content Understanding from a Collaborative Perspective framework (CURec), which generates collaborative-aligned content features for more comprehensive recommendations. 
CURec first adapts the LLM with recommendation objectives through post-training, equipping it with instruction-following and chain-of-thought reasoning capabilities. Next, we use the LLM-generated features to train the recommender model. Then, we treat the user–item prediction scores produced by the trained recommender model as reward signals to guide RL-based refinement of the LLM, thereby improving the quality of generated reasons and pattern. The corrected reasons are directly integrated into the recommender model to enhance comprehensibility and recommendation performance. Extensive experiments on public benchmarks demonstrate the superiority of CURec over existing methods.
\end{abstract}


\maketitle

\section{Introduction}
\label{sec:intro}

Recommender systems have become increasingly ubiquitous in daily life, powering applications such as e-commerce~\cite{LONGER}, short-video~\cite{PAM}, or social media~\cite{social} platforms.
Traditional recommendation algorithms~\cite{NCF, NGCF, SASRec} primarily rely on ID embedding features to represent users and items. While these approaches have demonstrated strong recommendation performance, they are inherently limited by their memorization-based nature~\cite{WaD}: the models tend to recall historical interactions while overlooking the semantic content information of items. To enrich the content representations of users and items, several methods~\cite{S3Rec, RecFORMER, UniSRec} incorporate additional item attributes, such as category labels or textual descriptions, into the recommendation process to enhance performance. Emerging approaches~\cite{LLMRec, KAR, LEARN} further leverage LLMs to generate features for recommendation, utilizing their reasoning capabilities to infer user behavior patterns~\cite{KAR} and their extensive world knowledge to uncover richer item information~\cite{LEARN}.

Enabling content understanding from the perspective of recommendation is critical for recommender systems~\cite{LEARN}, as it facilitates comprehensible recommendations, improves the efficiency of collaborative filtering, and enhances overall user experience. While incorporating various modalities of item information can alleviate the issue of limited content awareness, such approaches typically focus solely on the item's semantic information. They focus on introducing more precise descriptive information about items, rather than generating representations tailored to the recommendation objective. This neglect of the gap between semantic and collaborative perspectives results in a misalignment, preventing existing methods from achieving optimal recommendation performance.

\begin{figure}[t]
    \centering
    \includegraphics[width=\linewidth]{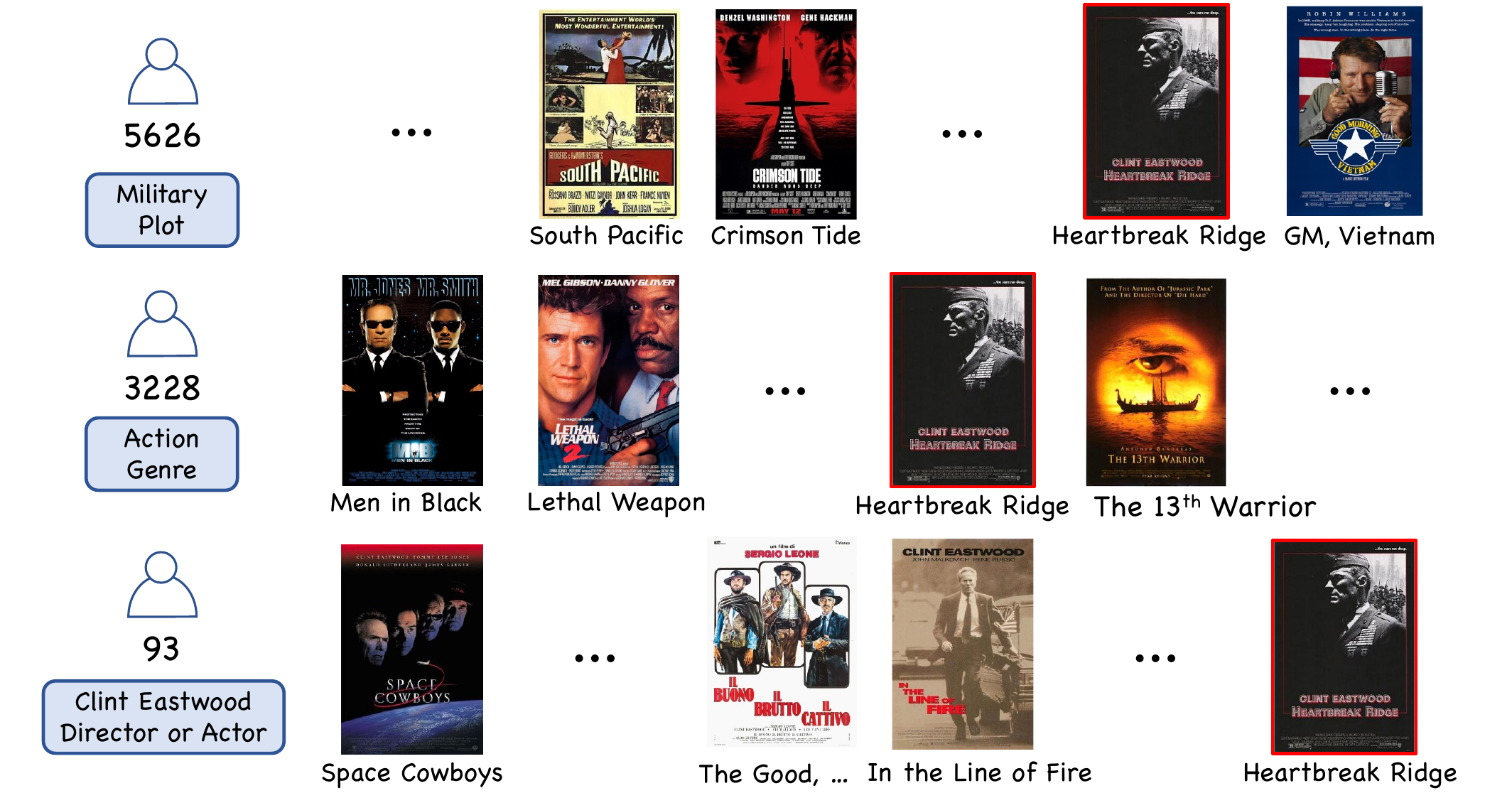}
    \caption{For the same movie, different users with distinct patterns (military plot, action genre, and Clint Eastwood as director or actor) may appreciate it for different reasons.}
    \label{fig:reasons}
\end{figure}

As illustrated in Fig.~\ref{fig:reasons}, different user interaction sequences reflect distinct interest patterns, which in turn lead to diverse reasons for liking the same movie.
These reasons from collaborative perspectives cannot be adequately described by semantic text information. However, such reasons are beneficial for generating comprehensible recommendations, as they provide deeper insights into user preferences and decision-making processes.

The rich world knowledge and powerful reasoning capabilities of LLMs~\cite{KAR, HiT} offer a promising solution for generating collaborative-perspective reasons by leveraging users' historical interaction sequences and the semantic descriptions of items. However, despite their impressive capabilities, directly applying LLMs to recommendation reason generation may yield suboptimal results~\cite{ReasoningRec}, such as incorrect, generic, or uninformative explanations. This is largely due to the misalignment between the training objectives of LLMs and the reasoning goals of recommender systems. Therefore, ensuring that LLMs are equipped with a recommendation-oriented chain-of-thought (CoT) reasoning process, as well as the ability to generate accurate and meaningful recommendation reasons, remains a significant challenge.

To this end, we propose a novel approach, Content Understanding from Collaborative Perspective framework (\method), which leverages LLMs to generate and refine recommendation-perspective content features and efficiently integrates them into recommender systems to enable content-aware, comprehensible recommendations. Specifically, \method consists of three key components: (1) LLM adaptation for recommender systems, 
(2) LLM-enhanced recommender training, and (3) recommender-guided LLM refinement.

Given the fundamental mismatch between the training objective of LLMs and recommender system models, to equip LLMs with the ability to follow structured instructions and understand the recommendation process through reasoning, we design the recommendation adaptation post-training for LLMs. The post-training is aimed at guiding the LLM to acquire the capability to leverage world knowledge and engage in chain-of-thought reasoning to complete recommendations for users. 
To better leverage the reasoning capabilities of LLM for improving recommendation performance and comprehensibility, we design a framework that utilizes post-trained LLM to generate user interest patterns and user–item recommendation reasons, and integrates these features into downstream recommendation tasks with an attention mechanism.
However, due to the hallucination problem inherent in LLMs~\cite{Hallucination}, directly applying the pretrained model to recommendation reason generation can lead to inaccurate or generic outputs.
Ideally, an end-to-end training paradigm would simultaneously enhance both the quality of LLM-generated reasons and the performance of the recommender model. However, in practice, there exists a substantial gap in data requirements and training time between LLM and lightweight recommender models, making end-to-end training inefficient.
To address this, we first use the LLM-generated features to train the recommender model. Then, we treat the user–item prediction scores produced by the trained recommender model as reward signals to guide RL-based refinement of the LLM, thereby improving the quality of generated reasons and patterns.
Finally, during online inference, by leveraging pre-stored LLM-generated features together with an activated lightweight recommender model, our approach achieves low inference latency. By explicitly modeling the attention-based matching between user interest patterns and item recommendation reasons lists, we achieve a more comprehensible and transparent recommendation process.

Our main contributions can be summarized as follows:

\begin{itemize}[leftmargin=*]
    \item We propose a chain-of-thought refinement framework based on RL, enabling LLMs to leverage world knowledge and perform recommendation-oriented reasoning, thereby generating accurate and personalized recommendation reasons.

    \item We introduce a model that effectively utilizes the generated user interest patterns and item recommendation reasons, achieving comprehensible and high-performance recommendation.

    \item Extensive experiments on public benchmark datasets demonstrate the effectiveness of the proposed \method framework.
\end{itemize}

\section{Related Works}

\subsection{Collaborative Filtering}

Collaborative filtering (CF)~\cite{CF} algorithms represent a milestone in recommendation systems, focusing on using historical user-item interaction data to predict potential future interactions. The main idea of CF is to generate representations for users and items and use their similarity to make recommendations~\cite{NCF}. In recent years, emerging approaches have gained prominence using GNNs~\cite{NGCF, LightGCN, GMCF, StarGCN, HCCF, SimRec}, aiming to treat user-item interactions as a bipartite graph, propagating information between users and items through the graph's links and optimizing the performance of similar nodes based on their interaction histories. Another common collaborative filtering approach models user representations based on their historical behavior sequences~\cite{Bert4Rec, SASRec, CASER, ICRec, DuoRec}, with many methods integrating GNNs for enhanced modeling~\cite{SRGNN, MAERec}. However, they rely solely on ID embeddings while ignoring rich content information, which limits their overall performance.

\subsection{Content-based Recommendation}

Content-based recommendation aims to alleviate the generalization limitations of recommender systems and enhance recommendation efficiency by incorporating side information about users and items. Early works, such as VBPR~\cite{VBPR}, optimize recommendation by combining visual features of items with ID embeddings. More recent approaches like M3CSR~\cite{M3CSR} and MAKE~\cite{MAKE} further improve the utilization of content information by introducing modality-specific encoders. Inspired by the BERT~\cite{BERT} encoding model, many studies~\cite{ZESRec, UniSRec, RecFORMER} integrate textual modalities to generate more informative item representations.
Although these methods effectively incorporate content features, they still fall short in enabling deep content understanding through reasoning. Moreover, they overlook the gap between the item-perspective and recommendation-perspective, resulting in suboptimal utilization of content information.

\subsection{LLM for Recommendation}

LLM-based recommendation methods can be broadly categorized into two types: using LLMs as recommender systems and leveraging LLMs to enhance recommendations. Approaches that employ LLMs directly as recommender systems~\cite{TALLRec, LlamaRec, CoLLM, BinLLM, LC-Rec, ReasoningRec, SeRALM} typically involve fine-tuning the language model using recommendation data or ID information.
However, using LLMs as recommenders faces practical limitations due to their high inference latency. Additionally, the mismatch between the training objectives of LLMs and recommender systems further hinders their ability to achieve optimal performance.
On the other hand, LLM-assisted recommendation~\cite{LLMRec, LEARN, KAR, HiT, AlphaRec, RLMRec} focuses on using LLMs to generate informative features.
Despite effectively incorporating knowledge and reasoning through LLMs, they have yet to address the challenge of content understanding from the recommendation perspective.

\section{Preliminary}
\subsection{Notations}
In this work, we consider the general CF scenario. Let $\mathcal{U}=\{u_1, \cdots,$ $ u_U\}$ and $\mathcal{I}=\{i_1, \cdots, i_I\}$ denote the universal set of users and items. $d_i$ stands for the text description of item $i$. $\mathcal{S}_u=\{i_1^u, \cdots, i_{|\mathcal{S}_u|}^u\}$ denotes the recent interaction sequence of user $u$, and $\mathcal{S}_i$ similarly denotes the interaction sequence of item $i$.
$(\mathcal{T}_u,\mathcal{T}_i,y_{ui})=((u, \mathcal{S}_u), (i,\mathcal{S}_i), y_{ui})\in\mathcal{D}$ construct an interaction sample in dataset $\mathcal{D}$, where $y_{ui}$ indicates the ground truth of interaction.

\subsection{Problem Formulation}
The objective of recommendation systems is to predict interaction labels as accurately as possible based on the existing information about users and items. For a given sample $(\mathcal{T}_u,\mathcal{T}_i, y_{ui})\in \mathcal{D}_\mathrm{train}$, the system takes the sample as inputs and outputs the predicted score of the interaction. The system is optimized as follows:
\begin{equation}
    \arg\min_{\Theta} \sum_{(\mathcal{T}_u,\mathcal{T}_i,y_{ui})\in \mathcal{D}_\mathrm{train}}\mathcal{L}\left(\mathcal{W}^{\Theta}(\mathcal{T}_u,\mathcal{T}_i), y_{ui}\right)
\end{equation}
where $\mathcal{W}$ represents the prediction recommender regarding to learnable parameters $\Theta$, and $\mathcal{L}$ denotes the loss function.

\section{Methods}

\subsection{Overview}

In this section, we will provide a detailed description of our proposed approach \method, as shown in Fig.~\ref{fig:framework}.

\textbf{LLM Adaptation for Recommender Systems.} First, we perform RL-based post-training of the LLM on recommendation tasks, enabling it to follow instructions in a specified format and conduct chain-of-thought reasoning. The adaptation also encourages the LLM to leverage world knowledge to perform reasoning adapted to recommendation objectives. Through this adaptation, the LLM gains a deeper understanding of the items in the system and the recommendation reasons behind why users might prefer them.

\textbf{LLM-Enhanced Recommender Training.} Next, to better leverage the reasoning capabilities of LLM for improving recommendation performance and comprehensibility, we utilize the post-trained LLM to generate user interest patterns and user–item recommendation reasons, and integrate these features into downstream recommendation tasks with an attention~\cite{attention} mechanism. In this way, when generating new reasons, they can be appended to the item's reason list and evaluated by the recommender model, allowing us to assess the quality of the generated explanations in a recommendation-aware manner.

\textbf{Recommender-Guided LLM Refinement.} After training the recommender model, we treat the user–item prediction scores produced by the model as reward signals to perform reinforcement learning-based refinement of the LLM, thereby improving the quality of generated reasons.
Finally, during inference, by leveraging pre-stored LLM-generated features together with an activated lightweight recommender model, our approach achieves inference latency comparable to that of traditional recommendation systems.

\begin{figure}[t]
    \centering
    \includegraphics[width=\linewidth]{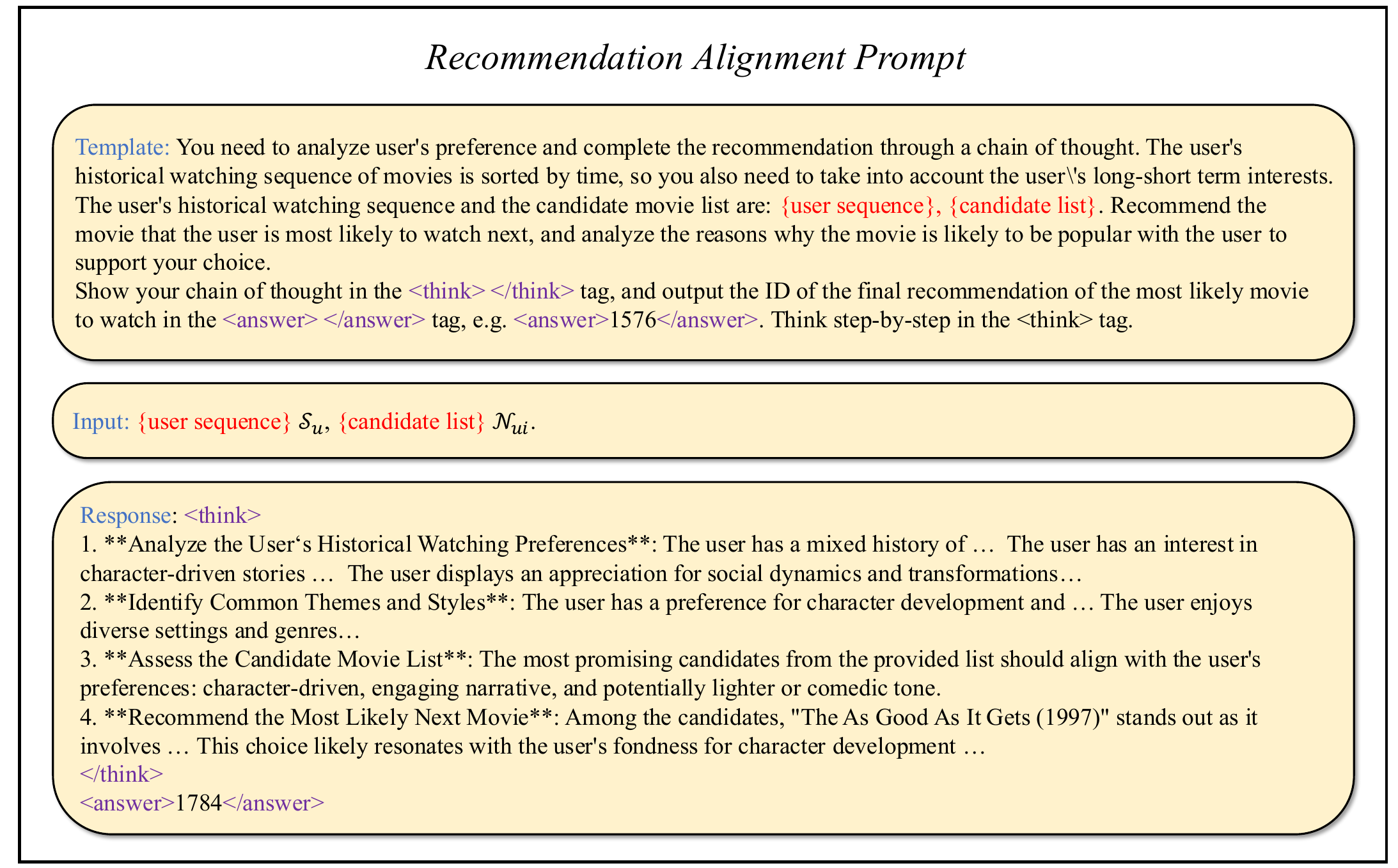}
    \caption{Example prompts for \textit{Recommendation Adaptation Prompt}, denoted as $prompt_{\text{rec}}$, which prompting the LLM to complete the recommendation task.}
    \label{fig:prompt_rec}
\end{figure}

\subsection{LLM Adaptation for Recommender Systems}

\begin{figure*}[t]
    \centering
    \includegraphics[width=\linewidth]{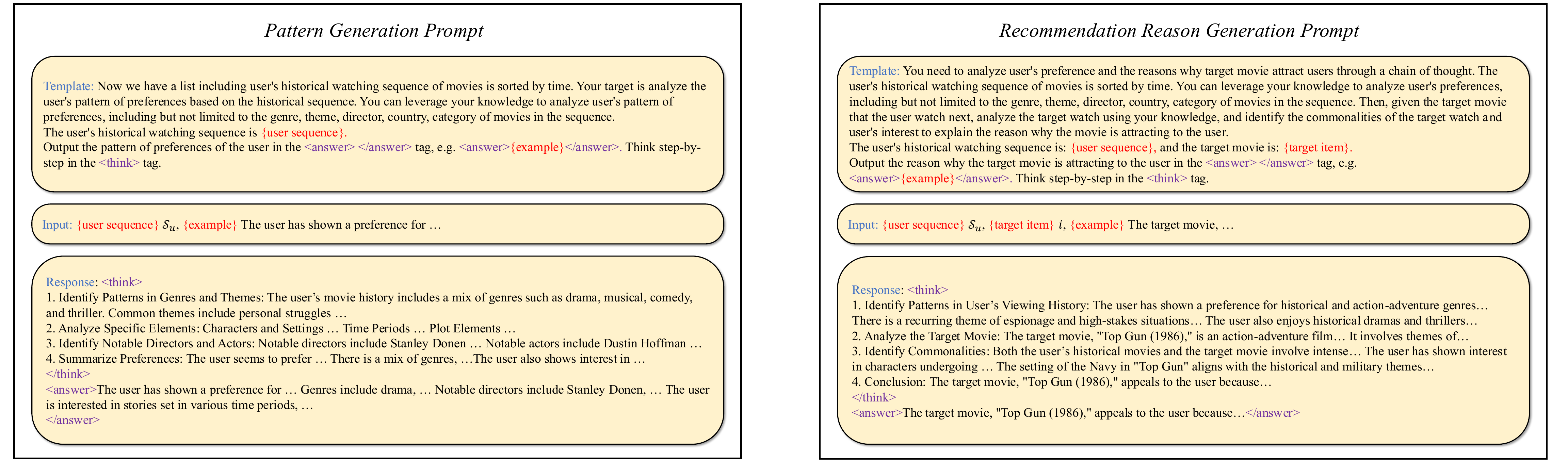}
    \caption{Example prompts for \textit{Pattern Generation Prompt} ($prompt_{\text{pattern}}$) and \textit{Recommendation Reason Generation Prompt} ($prompt_{\text{reason}}$), which prompting the LLM to generate patterns and recommendation reasons.}
    \label{fig:prompt_gen}
\end{figure*}

Since LLMs are primarily trained with objectives centered around dialogue completion and various domain-specific tasks (\textit{e.g.}, code, mathematics), directly applying them to recommendation tasks often yields suboptimal performance, despite their rich world knowledge and strong reasoning capabilities. Moreover, to effectively activate the reasoning abilities of LLMs, it is essential to guide them to produce outputs in a structured chain-of-thought format.

To bridge the gap between the training objectives of LLMs and recommender systems, we conduct recommendation adaptation post-training, as shown in Fig~\ref{fig:framework}(a). Specifically, we construct prompts based on recommendation data, prompting the LLM to perform CoT reasoning to complete the recommendation task. The generated completions are then evaluated via reward signals, which are used to fine-tune the LLM through reinforcement learning.

To address this, we design a \textit{Recommendation Adaptation Prompt} instruction template, denoted as $prompt_{\text{rec}}$, to guide the LLM in performing chain-of-thought reasoning, as illustrated in Fig.~\ref{fig:prompt_rec}. For each interaction sample $((u, \mathcal{S}_u), (i, \mathcal{S}_i))\in\mathcal{D}_{\text{train}}$, $prompt_{\text{rec}}$ instructs the model to analyze the descriptions of items in the user's interaction history $\mathcal{S}_u$, infer the user's interest pattern, and identify the most relevant item from a candidate list $\mathcal{N}_{ui}$. The candidate list $\mathcal{N}_{ui}$ consists of the target item $i$ from the interaction sample along with several negative items that do not appear in the user's interaction history $i^-, i\not\in\mathcal{S}_{u}$. We obtain the LLM's recommendation output by feeding it the constructed $prompt_{\text{rec}}$:

\begin{equation}
    o_{\text{rec}} = LLM(prompt_{\text{rec}}, \mathcal{S}_u, \mathcal{N}_{ui})
\end{equation}

To evaluate the outputs of the LLM, we design a rule-based reward $r$ to guide the RL training. The reward consists of three components:
(1) Format Reward: This checks whether the output follows the required structure, specifically whether the reasoning process is enclosed within <think> </think> tags and the answer within <answer> </answer> tags.
(2) Legal Reward: This checks whether the predicted item appears within the provided candidate list.
(3) Correctness Reward: This evaluates whether the predicted item matches the ground truth item from the interaction sample.

We train the LLM using GRPO~\cite{GRPO} based on the designed reward signals. For each $prompt_{\text{rec}}$, GRPO samples a group of responses $\{{o_{\text{rec}}}_1, {o_{\text{rec}}}_2, \cdots, {o_{\text{rec}}}_G\}$ and optimizes the policy model according to the following objective:

\begin{equation}
\begin{split}
    \mathcal{J}_{\text{GRPO}}(\theta, r) &= \mathbb{E}{[\{{o}_n\}_{n=1}^G \sim LLM_{\theta_{old}}(O|prompt, \mathcal{S}_u, \mathcal{N}_{ui})]}  \\
    & \frac{1}{G}\sum_{n=1}^G \left( \rho_n - \beta \mathbb{D}_{KL}\left(LLM_{\theta} || LLM_{ref}\right)\right), \\
    \rho_n &= \min \left[ \delta_n A_n, \text{clip} \left( \delta_n, 1 - \epsilon, 1 + \epsilon \right)  A_n \right], \\
    \delta_n &= \frac{LLM_\theta(o_n |prompt, \mathcal{S}_u, \mathcal{N}_{ui})}{LLM_{\theta_{old}}(o_n|prompt, \mathcal{S}_u, \mathcal{N}_{ui})}
\end{split}
\label{equ:GRPO-obj}
\end{equation}
where $\epsilon$ and $\beta$ are hyper-parameters, and $A_i$ is the advantage, computed using a group of rewards $\{{r}_1, {r}_2, \ldots, {r}_G\}$ corresponding to the outputs within each group:
\begin{equation}
    A_n = \frac{{r}_n - {\mathrm{mean}(\{{r}_1, {r}_2, \cdots, {r}_G\})}}{{\mathrm{std}(\{{r}_1, {r}_2, \cdots, {r}_G\})}}
    \label{equ:reward}
\end{equation}

Through the above RL-based training, we achieve an alignment between the LLM and the recommendation task. This enables the LLM to acquire both the reasoning ability required for recommendation objectives and the capability to follow structured output instructions. The pretrained LLM, denoted as $LLM$, becomes capable of effectively inferring user interest patterns and generating recommendation reasons in subsequent tasks.

\subsection{LLM-Enhanced Recommender Training}

\begin{figure*}[t]
    \centering
    \includegraphics[width=\linewidth]{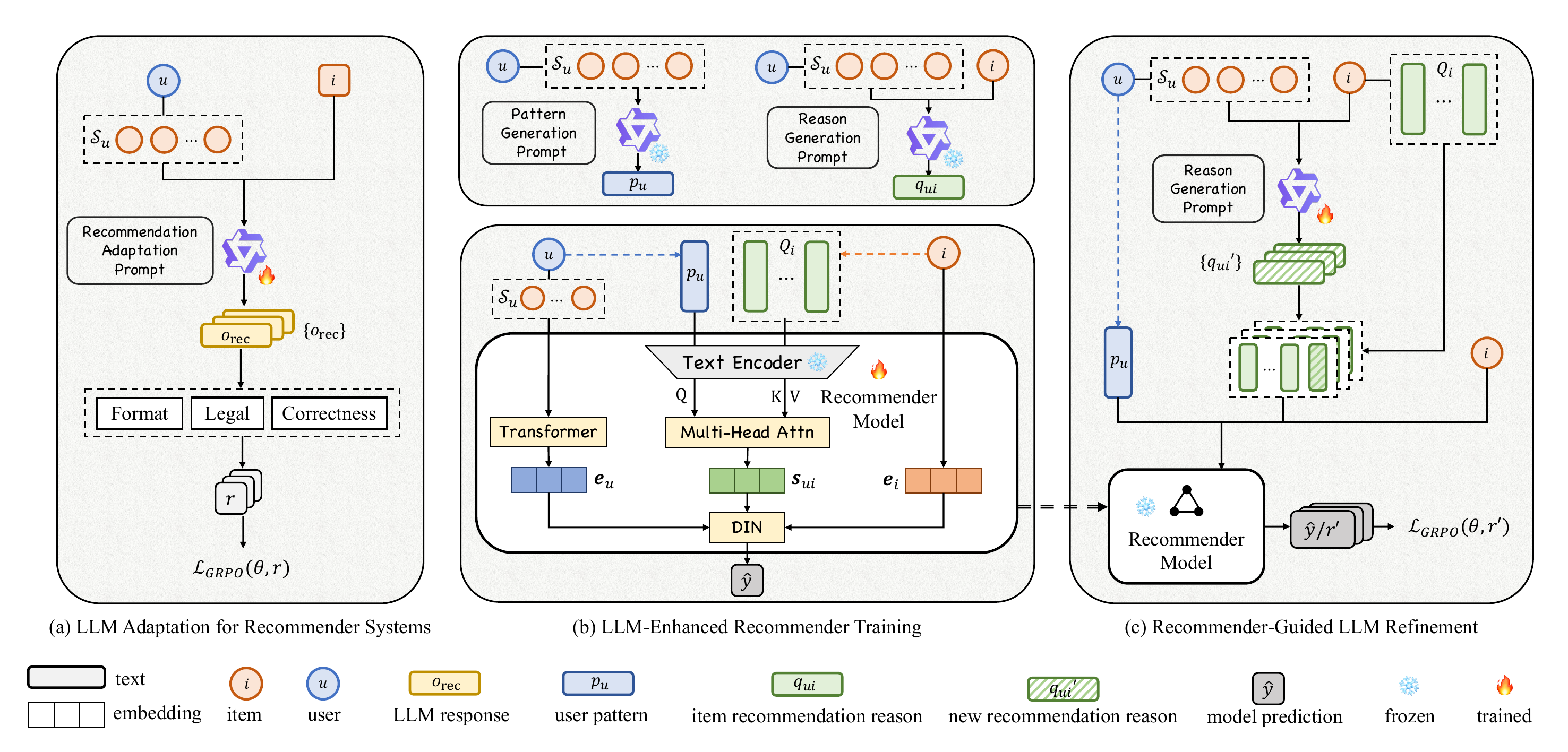}
    \caption{The framework of \method, including (a) LLM Adaptation for Recommender Systems, (b) LLM-Enhanced Recommender Training, and (c) Recommender-Guided LLM Refinement.}
    \label{fig:framework}
\end{figure*}

To better leverage the reasoning capabilities of LLM for improving recommendation performance and comprehensibility, we first utilize the post-trained LLM to generate user interest patterns and user–item recommendation reasons, and then integrate these features into recommendation tasks, as shown in Fig.~\ref{fig:framework}(b).

\subsubsection{Pattern and Reason Generation}
\label{sec:gen}

To extract user interest patterns and item recommendation reasons from interaction data using the pretrained LLM, we design two prompts: \textit{Pattern Generation Prompt} and \textit{Recommendation Reason Generation Prompt}, denoted as $prompt_{\text{pattern}}$ and $prompt_{\text{reason}}$, which guide the LLM to produce outputs through reasoning, as shown in Fig.~\ref{fig:prompt_gen}.

\textit{Pattern Generation Prompt} takes as input the descriptions of items from the user's interaction sequence and prompts the LLM to analyze the user's interests from multiple perspectives. For example, in the MovieLens dataset, $prompt_{\text{pattern}}$ directs the LLM to consider aspects including but not limited to genre, category, and director. For each user $u$, we use their interaction sequence $\mathcal{S}_u$ from the training set $\mathcal{D}_{\text{train}}$ as input and apply $prompt_{\text{pattern}}$ to obtain a summary of their personalized interest pattern:
\begin{equation}
    p_u = LLM(prompt_{\text{pattern}}, \mathcal{S}_u)
\end{equation}
where $p_u$ represents the generated pattern of the user.

\textit{Recommendation Reason Generation Prompt} additionally takes the description of the target item as input. $prompt_{\text{reason}}$ guides the LLM to infer why the user might favor the target item by identifying the commonalities between the user's interest pattern and the item's content description. The resulting reason is not a simple multi-perspective description of the item. Instead, it represents a recommendation-perspective content feature that integrates the LLM's rich world knowledge, the user's evolving interests, and the underlying reasons for user-item matching.
For each sample $((u, \mathcal{S}_u), (i, \mathcal{S}_i))$ in the training set $\mathcal{D}_{\text{train}}$, we use it as input to this prompt, allowing the LLM to generate a personalized reason explaining why the item is recommended to the user:
\begin{equation}
    q_{ui} = LLM(prompt_{\text{reason}}, \mathcal{S}_u, i)
    \label{equ:reason}
\end{equation}
where $q_{ui}$ denotes the generated recommendation reason. For each item, we aggregate all previously generated recommendation reasons into a reason list $Q_i = \{q_{ui}\}_{u\in\mathcal{S}_i}$, which serves as the item's content representation. This list provides a rich, recommendation-oriented understanding of the item from multiple angles. By doing so, we move beyond generic semantic descriptions and construct a content representation that reflects how different users perceive and relate to the item in the context of personalized recommendation.

\subsubsection{Recommender Training}
\label{sec:model}

For each user-item interaction sample, both the user interest pattern and the item recommendation reason are represented in a language semantic space. We encode them into embeddings using the LLM:

\begin{equation}
    \boldsymbol{e}_{p_u}=\mathrm{Encoder}(p_u), \boldsymbol{E}_{Q_i}=\{\mathrm{Encoder}(q)\}_{q\in Q_i}
\end{equation}

To compute the user-item matching representation, we apply a multi-head attention mechanism~\cite{attention}, where the user embedding serves as the query, and the list of item reason embeddings serves as the keys and values:
\begin{equation}
    \boldsymbol{s}_{ui} = \mathrm{MultiHeadAttention}(\boldsymbol{e}_{p_u}, \boldsymbol{E}_{Q_i}, \boldsymbol{E}_{Q_i})
\end{equation}
where $\boldsymbol{s}_{ui}$ represents the matching representation embedding.

This attention-based interaction allows the model to assess which recommendation reasons are most aligned with the user's interests. Through this approach, we can effectively evaluate whether specific aspects of the item resonate with the user's preferences, thereby enhancing the comprehensibility of the recommendation process.

To effectively train the attention parameters and derive user-item matching scores that serve as reward signals, we integrate the matching representation into a traditional recommendation model and train it using recommendation data. Following common practices in sequential recommendation~\cite{SASRec}, we encode the user's interaction history as a sequence of item embeddings and process it through a Transformer~\cite{attention} to obtain a user embedding:
\begin{equation}
    \boldsymbol{e}_u=\boldsymbol{h}_1, \{\boldsymbol{h}_i\}_{i\in\mathcal{S}_u}=\mathrm{Transformer}\left(\{\boldsymbol{e}_i\}_{i\in\mathcal{S}_u}\right)
\end{equation}
where $\boldsymbol{e}_i$ is the item ID embedding. Next, we compute the prediction score by combining the user and item embedding, and the matching representation using the Deep Interest Network (DIN)~\cite{DIN}:

\begin{equation}
    \hat{y}_{ui}=\mathrm{DIN}([\boldsymbol{e}_u, \boldsymbol{e}_i, \boldsymbol{s}_{ui}])
\end{equation}
where $\hat{y}_{ui}$ is the prediction score. Finally, we train the model using the negative log-likelihood loss:
\begin{equation}
    \mathcal{L} = -\log\frac{\exp(\hat{y}_{ui})}{\sum_{i^-\in \mathcal{I}}\exp(\hat{y}_{ui^-})}
\end{equation}

\subsection{Recommender-Guided LLM Refinement}

Although the post-trained LLM can effectively generate user interest patterns and item recommendation reasons through reasoning, the quality of its outputs is not always reliable. Issues such as hallucinations can lead to incorrect or overly generic reasons. To address this issue, we innovatively use the prediction scores from the recommendation model as reward signals to refine the LLM, thereby optimizing the quality of the generated recommendation reason features,
as illustrated in Fig.~\ref{fig:framework}(c).

Similar to the adaptation post-training stage, we fine-tune the LLM using a RL approach. We adopt the $prompt_{\text{reason}}$ described in Sec.~\ref{sec:gen} to guide the LLM in reasoning and generating a new recommendation reason $q_{ui}'$, following the formulation in Equ.~(\ref{equ:reason}). The newly generated reason is appended to the item's reason list and evaluated using the recommender model:

\begin{equation}
    r' = \hat{y}_{ui} = f(p_u, Q_i\cup\{q_{ui}'\},\mathcal{S}_u, i)
\end{equation}
where $f$ represent the recommender model described in Sec.~\ref{sec:model}.

We adopt the GRPO training framework. For each interaction, we sample a group of recommendation reasons, freeze the recommender model and compute the corresponding reward signals. Based on Equ.~(\ref{equ:GRPO-obj}-\ref{equ:reward}), we then calculate the loss and perform a one-step fine-tuning update on the LLM:
\begin{equation}
    \theta_{new} = \theta_{old} - \alpha\nabla_{\theta}\mathcal{J}_{\mathrm{GRPO}}(\theta,r')
\end{equation}
where $\theta$ refer to the parameter of LLM, and $\alpha$ is the learning rate.

Through this training paradigm, we effectively improve both the comprehensibility and transparency of the recommendation process, while simultaneously enhancing the quality of recommendation reasons generated by the LLM. During online inference, the pattern and reason features generated by the fine-tuned LLM can be precomputed and stored, and then fed into the lightweight recommendation model to produce predictions. This design enables low inference overhead comparable to traditional recommenders.

\begin{table}[t]
    \caption{Overview of the datasets.}
    \label{tab:dataset}
    {\small
    \begin{tabular}{c|cccc}
    \toprule
    \textbf{Dataset} & \textbf{\#Users} & \textbf{\#Items} & \textbf{\#Inter.}&\textbf{Density}\\ \midrule
    \textbf{MovieLens}& 6,041& 3,952& 575,292&2.41\%\\
    \textbf{Video Games}& 15,459& 12,540& 227,575&1.17\textperthousand\\
    \textbf{Movies and TV}& 12,722& 20,551& 405,189&1.55\textperthousand\\\bottomrule
    \end{tabular}
    }
    \vspace{-3mm}
\end{table}

\section{Experiments}
We conducted experiments to answer the following RQs: 

\textbf{RQ1}: How does \method perform compared to current baselines?

\textbf{RQ2}: How much of a performance enhancement do various components of \method provide? 

\textbf{RQ3}: How does \method compare to other baseline models in terms of inference efficiency?

\textbf{RQ4}: Can \method generate high-quality user patterns and recommendation reasons to enhance recommendation comprehensibility?

\subsection{Experimental Settings}

\subsubsection{Datasets}
We selected three public datasets that contain timestamps, including \textbf{MovieLens}, \textbf{Video Games}, and \textbf{Movie and TVs}. For the label processing of the datasets, the user ratings lie between 0 and 5, while we consider the interaction data with ratings above 3 as positive samples and remove all negative samples. We retained users with 30 and items with 10 or more interactions. The information on datasets is presented in Table~\ref{tab:dataset}.

\textbf{MovieLens}-1M~\cite{MovieLens} is a dataset of user ratings for movies.\textbf{Video Games} and \textbf{Movies and TV} are a part of the \textbf{Amazon-2023}~\cite{Amazon} dataset, which collects games, movies and TV reviews from purchasers on the Amazon e-shopping site.

\begin{table*}[t]
\centering
\caption{The reported top-K evaluation metrics. Bold represents the optimal and underlined represents the suboptimal results.}
\label{tab:perf}
\resizebox{\textwidth}{!}{%
{\small
\begin{tabular}{c|c|cccccccccccc|c}
\toprule
\textbf{Dataset} & Metric & NCF & SASRec&Bert4Rec & DuoRec& MAERec& UniSRec&RFormer&LLMRec& KAR&LEARN &AlphaRec& \method & Impr. \% \\ \midrule
\multirow{7}{*}{\textbf{MovieLens}} 
& Recall@1 &0.0101& 0.0222&0.0286& 0.0338& 0.0365& 0.0346& 0.0375&0.0339& \underline{0.0393}&0.0297 &0.0365&\textbf{0.0447}& +13.74\%\\
\cmidrule{2-15}
& Recall@5 &0.0370& 0.1012&0.1029& 0.1143& 0.1166& 0.1124& 0.1178&0.1205& \underline{0.1225}&0.1016 &0.1215&\textbf{0.1457}& +18.94\%\\
& Recall@10 &0.0539& 0.1687&0.1645& 0.1766& 0.1877& 0.1813& 0.1831&0.1867& \underline{0.1914}&0.1810 &0.1913&\textbf{0.2195}& +14.68\%\\
& Recall@20 &0.0820& 0.2639&0.2545& 0.2503& 0.2721& 0.2728& 0.2642&0.2778& 0.2787&0.2603 &\underline{0.2807}&\textbf{0.3096}& +10.29\%\\
\cmidrule{2-15}
&NDCG@5 &0.0233& 0.0616&0.0656& 0.0742& 0.0770& 0.0738& 0.0761&0.0778& \underline{0.0807}&0.0681 &0.0794&\textbf{0.0952}& +17.97\%\\
&NDCG@10 &0.0288& 0.0834&0.0854& 0.0944& 0.0998& 0.0960& 0.1003&0.0992& \underline{0.1029}&0.0909 &0.1015&\textbf{0.1192}& +15.84\%\\
&NDCG@20 &0.0358& 0.1073&0.1082& 0.1128& \underline{0.1254}& 0.1190& 0.1233&0.1221& 0.1249&0.1098 &0.1244&\textbf{0.1419}& +13.16\%\\
\midrule
\multirow{7}{*}{\textbf{Video Games}}
& Recall@1 &0.0031& 0.0089&0.0130& 0.0194& 0.0131& 0.0121& 0.0146&0.0130& \underline{0.0182}&0.0124 &0.0153&\textbf{0.0212}& +16.48\%\\
\cmidrule{2-15}
& Recall@5 &0.0104& 0.0354&0.0424& 0.0501& 0.0478& 0.0404& \underline{0.0502}&0.0426& 0.0484&0.0405 &0.0473&\textbf{0.0587}& +16.93\%\\
& Recall@10 &0.0186& 0.0636&0.0686& 0.0748& \underline{0.0802}& 0.0668& 0.0747&0.0670& 0.0750&0.0610 &0.0741&\textbf{0.0896}& +11.72\%\\
& Recall@20 &0.0322& 0.1066&0.1070& 0.1083& \underline{0.1275}& 0.1066& 0.1152&0.1015& 0.1085&0.0911 &0.1098&\textbf{0.1330}& +4.31\%\\
\cmidrule{2-15}
&NDCG@5 &0.0067& 0.0220&0.0276& \underline{0.0351}& 0.0300& 0.0262& 0.0323&0.0276& 0.0334&0.0249 &0.0314&\textbf{0.0403}& +14.81\%\\
&NDCG@10 &0.0093& 0.0311&0.0360& \underline{0.0430}& 0.0404& 0.0347& 0.0402&0.0276& 0.0420&0.0316 &0.0401&\textbf{0.0502}& +16.74\%\\
&NDCG@20 &0.0127& 0.0419&0.0457& 0.0515& \underline{0.0523}& 0.0448& 0.0504&0.0441& 0.0504&0.0418 &0.0491&\textbf{0.0610}& +16.63\%\\
\midrule
\multirow{7}{*}{\textbf{Movies and TV}}
& Recall@1 &0.0033& 0.0189&0.0289& 0.0290& 0.0236& 0.0249& 0.0326&0.0404& 0.0406&0.0341 &\underline{0.0410}&\textbf{0.0492}& +20.00\%\\
\cmidrule{2-15}
& Recall@5 &0.0104& 0.0428&0.0494& 0.0499& 0.0454& 0.0512& 0.0547&0.0614& 0.0598&0.0592 &\underline{0.0626}&\textbf{0.0721}& +15.18\%\\
& Recall@10 &0.0187& 0.0599&0.0641& 0.0654& 0.0757& 0.0679& 0.0713&0.0744& 0.0758&0.0722 &\underline{0.0766}&\textbf{0.0906}& +19.53\%\\
& Recall@20 &0.0319& 0.0865&0.0850& 0.0904& 0.0911& 0.0951& 0.0921&0.0944&0.0951&0.0911 &\underline{0.0965}&\textbf{0.1153}& +18.28\%\\
\cmidrule{2-15}
&NDCG@5 &0.0068& 0.0309&0.0392& 0.0398& 0.0347& 0.0382& 0.0434&0.0512& 0.0504&0.0422 &\underline{0.0521}&\textbf{0.0602}& +15.55\%\\
&NDCG@10 &0.0095& 0.0365&0.0439& 0.0448& 0.0400& 0.0437& 0.0488& 0.0554&0.0556&0.0471 &\underline{0.0566}&\textbf{0.0661}& +16.78\%\\
&NDCG@20 &0.0128& 0.0432&0.0492& 0.0510& 0.0473& 0.0505& 0.0540&0.0603&0.0604&0.0578 &\underline{0.0616}&\textbf{0.0723}& +17.37\%\\
\bottomrule
\end{tabular}}
}
\end{table*}

\subsubsection{Baselines}
We compare \method with the following 11 baselines from diffrernt groups.

\textit{(i)} Traditional collaborative filtering:

- \textbf{NCF}~\cite{NCF} is one of the early methods that introduced deep learning into recommendation systems, leveraging MLP to model the similarity between users and items. 

- \textbf{SASRec}~\cite{SASRec} utilize a unidirectional Transformer to model item sequences to generate user representations.

- \textbf{Bert4Rec}~\cite{Bert4Rec} utilizes a bi-directional Transformer as the backbone to model the historical sequence of users.

- \textbf{DuoRec}~\cite{DuoRec} improves modeling performance through model-level expansion and contrastive normalization. 

- \textbf{MAERec}~\cite{MAERec} designs a task-adaptive augmentation mechanism based on GNN to enhance representation.

\textit{(ii)} Content-based recommendation:

- \textbf{UniSrec}~\cite{UniSRec} uses textual item representations from a pretrained language model and adapts to a new domain using an MoE-enhance adaptor.

- \textbf{RecFormer}~\cite{RecFORMER} replaces traditional item ID sequences with sequences of words from item textual descriptions and employs a bidirectional attention mechanism to encode them.

\textit{(iii)} LLM enhanced recommendation:

- \textbf{LLMRec}~\cite{LLMRec} estimate potential interactions and generates user and item attributes using LLMs. 

- \textbf{KAR}~\cite{KAR} leverages LLMs to incorporate external knowledge into the recommendation process by encoding user interests and item descriptions as additional features.

- \textbf{LEARN}~\cite{LEARN} employs LLMs to encode item embeddings from textual descriptions, then builds a dual-tower architecture using Transformer networks.

- \textbf{AlphaRec}~\cite{AlphaRec} introduces a trainable MLP to transform item representations produced by large language models into collaborative representations.

\subsubsection{Evaluation Metrics}
In our setup, we utilize the widely adopted Recall and NDCG~\cite{NDCG} metrics to evaluate the performance of various recommendation models. Following prior work, we adopt the commonly used leave-one-out~\cite{SASRec} evaluation strategy. Specifically, for each user, we hold out the most recent interaction as the test instance and use the remaining interactions for training. During evaluation, we perform a full ranking over the entire item set, rather than relying on negative sampling. Finally, we report the average values of the metrics across all samples in the test set.

\subsubsection{Implementation Details}
For the RL fine-tuning of the LLM during both the adaptation and refinement phases, we use Qwen2.5-7B~\cite{Qwen} as the base model. We use the first 80\% of each user’s interaction sequence for the post-training stage, and the remaining 20\% for the refinement phase.
The training is conducted using the AdamW~\cite{AdamW} optimizer with a learning rate of 5e-7, and we apply a cosine learning rate scheduler. We train the model for 2 epochs on each dataset to ensure effective alignment without overfitting.
During the training process of the recommender model, we utilize Adam~\cite{Adam} as the optimizer with the learning rate set to 1e-3. The batch size is set to 128, and the embedding dimension is 64. Codes are available at \url{https://github.com/Sycamoretail/CURec}.

\subsubsection{Training Cost}
For the post-training of the LLM with RL on each dataset, we adopt a multi-node training setup, where each training run uses two machines, each equipped with four NVIDIA L20 GPUs with 48GB memory. On average across all datasets, the recommendation-aligned post-training consumed 79.6 GPU-hours, while the CoT refinement stage consumed 27.7 GPU-hours.

\subsection{Performance Comparison (RQ1)}

\subsubsection{Overall Performance}

Tab.~\ref{tab:perf} presents the top-K evaluation results of various methods across datasets, where K is set to 1, 5, 10, and 20. Based on the results, we make the following observations:

First, among the three groups of baselines, we observe a clear performance hierarchy. Traditional CF methods (\textit{e.g.}, DuoRec, MAERec) achieve reasonable results through advanced sequence modeling and augmentation techniques, yet remain limited by their reliance on ID-based representations without external knowledge. Content-based methods (UniSRec, RecFormer) narrow the gap by incorporating textual item features; notably, on movie-related datasets where content descriptions are richer and more informative, these methods outperform several CF baselines. However, their performance is constrained by the lack of recommendation-oriented reasoning—they describe items from a semantic perspective rather than from a collaborative one. LLM-enhanced methods (KAR, AlphaRec) further improve upon content-based approaches by introducing LLM-generated knowledge, achieving the strongest baseline results across most metrics.
We also note that relying solely on frozen content features (UniSRec, LEARN) without trainable ID-based representations often leads to suboptimal performance, indicating that collaborative signals and content features are complementary.

Second, our proposed \method consistently outperforms all baselines by a substantial margin across all datasets and metrics, with average relative improvements of 14.95\%, 13.95\%, and 17.53\% in Recall@5 on MovieLens, Video Games, and Movies and TV, respectively. Compared to other LLM-enhanced methods, \method offers two key advantages: \textit{(i)} it leverages chain-of-thought reasoning to generate recommendation-adapted features rather than generic item descriptions, and \textit{(ii)} it employs recommender-guided reinforcement learning to refine the LLM, producing more accurate and personalized content features. These results demonstrate the necessity of bridging the semantic-collaborative gap through recommendation-oriented reasoning and iterative refinement.

\subsubsection{Cold-start User Performance}

\begin{table}[t]
    \centering
    \caption{Cold-start user performance.}
    {\small
    \resizebox{\columnwidth}{!}{%
    \begin{tabular}{c|c|ccc|ccc}
        \toprule
        \multicolumn{2}{c|}{\textbf{Dataset}} & \multicolumn{3}{c|}{\textbf{MovieLens}}& \multicolumn{3}{c}{\textbf{Video Games}}\\
        \midrule \multicolumn{2}{c|}{Metric} & R@1& R@5 & N@5  & R@1& R@5 & N@5\\ 
        \midrule
        \multicolumn{2}{c|}{Bert4Rec} & 0.0282& 0.1090& 0.0735 & 0.0110& 0.0391& 0.0263\\
        \multicolumn{2}{c|}{MAERec} & 0.0361& 0.1116& 0.0752 & 0.0122& 0.0461& 0.0299\\
        \multicolumn{2}{c|}{KAR} & 0.0360& 0.1160& 0.0761 & 0.0144& 0.0426& 0.0288\\
        \multicolumn{2}{c|}{AlphaRec} & 0.0348& 0.1182& 0.0768 & 0.0145& 0.0465& 0.0306\\
        \multicolumn{2}{c|}{\method} & \textbf{0.0410}& \textbf{0.1297}& \textbf{0.0854} & \textbf{0.0206}& \textbf{0.0527}& \textbf{0.0368}\\
        \bottomrule
    \end{tabular}
    }
    \label{tab:cold-user}
    }
\end{table}

\begin{table}[t]
    \centering
    \caption{Cold-start item performance.}
    {\small
    \resizebox{\columnwidth}{!}{%
    \begin{tabular}{c|c|ccc|ccc}
        \toprule
        \multicolumn{2}{c|}{\textbf{Dataset}} & \multicolumn{3}{c|}{\textbf{MovieLens}}& \multicolumn{3}{c}{\textbf{Video Games}}\\
        \midrule \multicolumn{2}{c|}{Metric} & R@1& R@5 & N@5  & R@1& R@5 & N@5\\ 
        \midrule
        \multicolumn{2}{c|}{Bert4Rec} & 0.0027& 0.0131& 0.0066 & 0.0057& 0.0125& 0.0088\\
        \multicolumn{2}{c|}{MAERec} & 0.0056& 0.0131& 0.0094 & 0.0049& 0.0130& 0.0089\\
        \multicolumn{2}{c|}{KAR} & 0.0094& 0.0244& 0.0175 & 0.0063& 0.0123& 0.0094\\
        \multicolumn{2}{c|}{AlphaRec} & 0.0092& 0.0259& 0.0181 & 0.0064& 0.0130& 0.0096\\
        \multicolumn{2}{c|}{\method} & \textbf{0.0169}& \textbf{0.0281}& \textbf{0.0227} & \textbf{0.0079}& \textbf{0.0144}& \textbf{0.0106}\\
        \bottomrule
    \end{tabular}
    }
    \label{tab:cold-item}
    }
\end{table}

We further evaluate the performance of different models on cold-start users. Specifically, we first count the number of interactions for each user in the training set, and then select the bottom 50\% of users in terms of interaction frequency as cold-start users. We retain the test results of these users as the cold-start evaluation set, as reported in Tab.~\ref{tab:cold-user}. The results show that \method outperforms all baseline models, which can be attributed to the recommendation-adapted post-training that enables the LLM to effectively extract user preference patterns from limited interactions, thereby improving recommendation performance.

\subsubsection{Cold-start Item Performance}

We also analyze the performance on cold-start items. Similarly, we select the bottom 50\% of items in terms of interaction frequency and retain their corresponding instances in the test set for evaluation, with the results reported in Tab.~\ref{tab:cold-item}. It can be observed that, although the limited number of interactions leads to fewer generated recommendation reasons for these items, \method still outperforms other baselines. This is because the LLM, after recommendation-adapted post-training and chain-of-thought refinement, is able to analyze items from multiple perspectives and generate high-quality collaborative-oriented textual features, which in turn improve performance in downstream comprehensible recommendation tasks.

\subsection{Ablation Study (RQ2)}
In this subsection, we conduct ablation studies on the proposed \method to analyze the effectiveness of its components. The experimental results are presented in Tab.~\ref{tab:ablation}. 

\begin{table}[t]
    \centering
    \caption{Ablation study of key components of \method.}
    {\small
    \resizebox{\columnwidth}{!}{%
    \begin{tabular}{c|c|cc|cc|cc}
        \toprule
        \multicolumn{2}{c|}{\textbf{Dataset}} & \multicolumn{2}{c|}{\textbf{MovieLens}} & \multicolumn{2}{c|}{\textbf{Video Games}} & \multicolumn{2}{c}{\textbf{Movies and TV}} \\
        \midrule \multicolumn{2}{c|}{Metric} & R@5 & N@5 & R@5 & N@5& R@5 & N@5 \\ 
        \midrule
        \multicolumn{2}{c|}{Text} & 0.1203& 0.0785& 0.0495& 0.0333& 0.0625&0.0509\\
        \midrule
        \multicolumn{2}{c|}{\texttt{w/o} Refine}  & 0.1353& 0.0890& 0.0505& 0.0346& 0.0704& 0.0576\\
        \midrule
        \multirow{3}{*}{\makecell{Diff. \\ Util.}}
        & AP & 0.1368& 0.0909& 0.0539& 0.0367& 0.0670& 0.0563\\
        & MLP & 0.1388& 0.0911& 0.0528& 0.0357& 0.0638& 0.0536\\
        & SA & 0.1397& 0.0923& 0.0541& 0.0363& 0.0659& 0.0557\\
        \midrule
        \multirow{2}{*}{\makecell{Diff. \\ Fus.}}
        & MLP & 0.1425& 0.0921& 0.0519& 0.0355& 0.0683& 0.0578\\
        & BI & 0.1434& 0.0936& 0.0562& 0.0384& 0.0713& 0.0598\\
        \midrule
        \multicolumn{2}{c|}{\method} & \textbf{0.1457}& \textbf{0.0952}& \textbf{0.0587}& \textbf{0.0403}& \textbf{0.0721}& \textbf{0.0602}\\
        \bottomrule
    \end{tabular}
    }
    \label{tab:ablation}
    }
\end{table}

\subsubsection{Effect of Introducing Reasons} 

In this section, we analyze the impact of introducing collaborative-perspective recommendation reasons. When reasons are entirely excluded and replaced with raw item textual descriptions, encoded and used alongside user patterns as features within a traditional recommender (variant Text), the model performs similarly to other LLM-assisted baseline methods. This result highlights a key insight: simply providing more detailed descriptions of item content is insufficient for optimal performance. Instead, incorporating content understanding from a collaborative perspective is more effective in supporting and enhancing recommendation comprehensibility.

\subsubsection{Effect of LLM Refinement}

In this section, we analyze the impact of the recommender-guided LLM refinement. The results show that omitting the refinement process (\texttt{w/o} Refine) leads to a significant performance drop, indicating that the refinement process effectively enhances the LLM's ability to generate accurate recommendation reasons and updates content features accordingly.
From another perspective, even without refinement, the model still outperforms most baseline methods, demonstrating the effectiveness of generating collaborative-perspective content features from an LLM to enhance recommendation performance.

\subsubsection{Effect of Reasons Utilization}

In this section, we analyze the impact of different strategies for utilizing recommendation reasons. We compare three integration methods for the item reason list: average pooling (AP), MLP, and self-attention (SA). In each case, the fused reason representation is concatenated with the user interest pattern and replaces the attention-based matching representation in \method. The results show that applying an attention mechanism to integrate the user interest pattern with the item reason list yields the best performance. This approach not only enhances recommendation performance but also improves comprehensibility, as it explicitly models the alignment between user interests and specific recommendation reasons.

\subsubsection{Effect of Fusion model}

In this section, we evaluate the impact of different feature fusion strategies. Specifically, we replace the DIN-based fusion with concat\&MLP as well as bilinear fusion (BI). The results indicate that the fusion method has only a minor impact on performance, because the user pattern and item reason feature inputs are of high quality and the attention-based matching is already sufficiently effective.

\subsubsection{Effect of LLM Scale}

\begin{figure}[t]
    \centering
    \includegraphics[width=.6\linewidth]{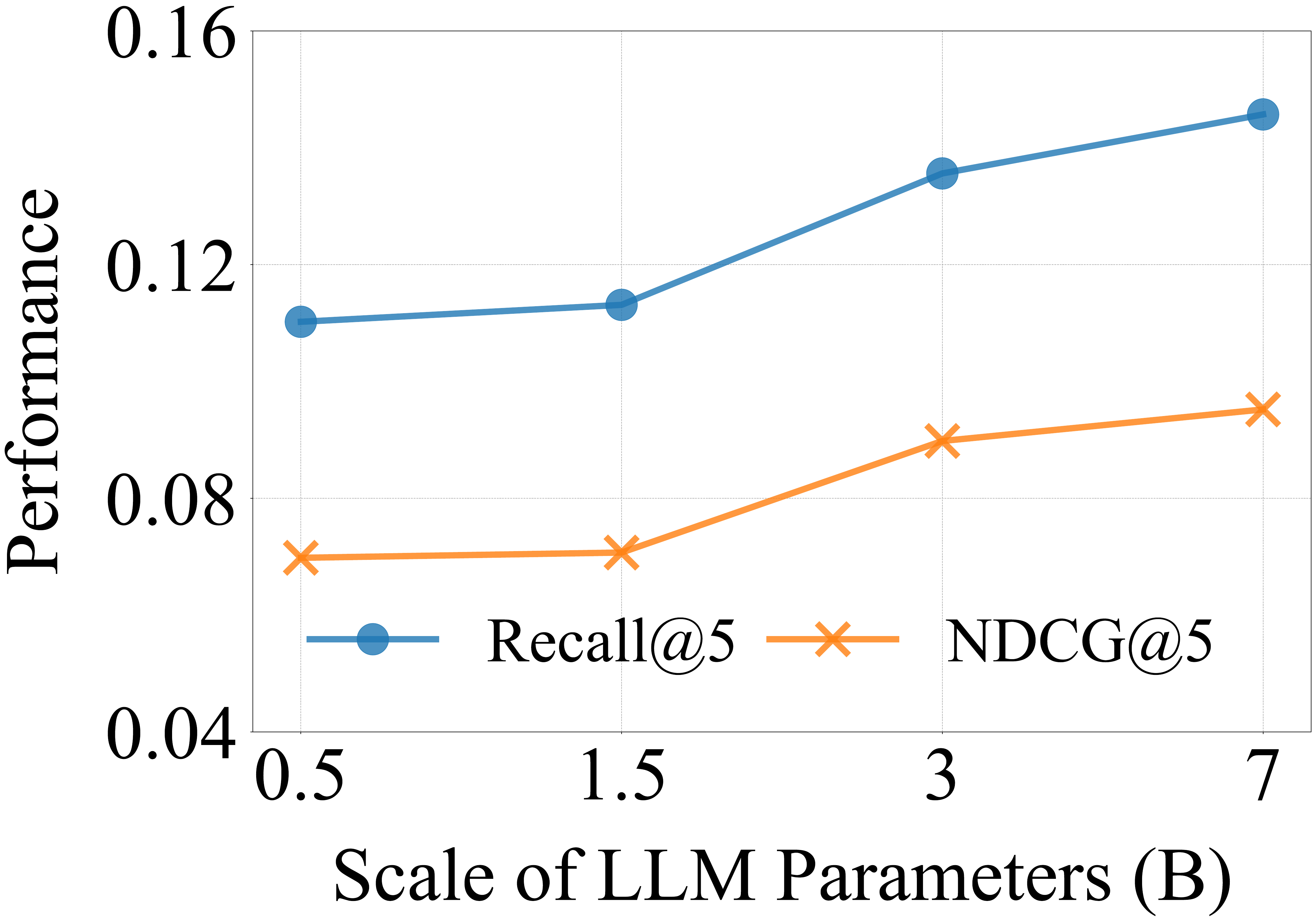}
    \caption{The performance of \method \textit{w.r.t.} LLM scale.}
    \label{fig:scale}
\end{figure}

In this section, we further analyze the impact of LLM scale on reason generation quality, as shown in Fig.~\ref{fig:scale}. The results indicate that as the number of LLM parameters increases, performance improves correspondingly, reflecting the generation of higher-quality recommendation reasons.
Notably, we observe a significant performance drop when using models with fewer than 3B parameters. This is because models below this threshold lack sufficient reasoning capabilities~\cite{reasoning}.

\subsection{Efficiency Analysis (RQ3)}

\begin{table}[t]
    \centering
    \caption{The comparison of inference time (s).}
    \begin{tabular}{ccc}
    \toprule
    \textbf{Dataset} & \textbf{MovieLens}& \textbf{Video Games}\\
    \midrule
    Qwen2.5 API& $7.421$ & $7.228$ \\
    Qwen2.5 \texttt{w/} Adapt&  $1.013$ & $1.052$ \\
    MAERec& $2.965\times 10^{-3}$ & $2.943\times 10^{-3}$\\
    SASRec& $2.889\times 10^{-3}$ & $2.796\times 10^{-3}$\\
    \method& $2.941\times 10^{-3}$ & $2.942\times 10^{-3}$ \\
    \bottomrule
    \end{tabular}
    \label{tab:inference}
\end{table}

In this section, we analyze the inference time efficiency of different methods. Qwen2.5 API refers to the setting where user history, negative samples, and item descriptions are provided as prompts to the raw LLM to directly perform the recommendation task. Qwen2.5 \texttt{w/} Adapt represents LLM after recommendation-adapted post-training in \method, where the LLM is fine-tuned to follow instruction formats and reason effectively for recommendation tasks. The remaining methods are based on traditional recommendation models.

In the evaluation of traditional recommendation models, we use a batch size of 128 during inference. For LLM-based methods, we adopt a smaller batch size of 10, due to GPU memory constraints. We evaluate the average inference time per sample for all methods in Tab.~\ref{tab:inference}.
The results demonstrate that directly applying LLMs to perform recommendation tasks incurs extremely high latency, which is unacceptable for real-time applications with strict timeliness requirements.
In contrast, \method achieves lower inference latency by offloading the LLM-based reasoning to a lightweight recommendation model. Compared to SASRec, \method introduces only an acceptable increase in latency, while delivering superior performance.
Moreover, \method outperforms complex models such as MAERec in terms of inference efficiency.

\subsection{Comprehensibility Analysis (RQ4)}
Beyond recommendation accuracy, a central goal of \method is to enhance the comprehensibility of the recommendation process—\textit{i.e.}, providing human-interpretable explanations for why an item is recommended to a specific user. In this section, we evaluate the quality of LLM-generated recommendation reasons through both automatic evaluation and a qualitative case study.

\subsubsection{Improvements on Comprehensibility}

\begin{table}[t]
    \centering
    \caption{The GSB test result for post-training and refinement.}
    \begin{tabular}{ccc}
    \toprule
    \textbf{Dataset} & \textbf{MovieLens}& \textbf{Video Games}\\
    \midrule
    After Post-training& 61.0\%/18.5\%/20.5\% & 59.0\%/19.0\%/22.0\% \\
    After Refinement& 49.5\%/25.5\%/25.0\% &  46.0\%/28.5\%/25.5\% \\
    \bottomrule
    \end{tabular}
    \label{tab:GSB}
\end{table}

To systematically evaluate whether the proposed training pipeline improves the comprehensibility of LLM-generated recommendation reasons, we adopt an LLM-as-a-judge paradigm and employ GPT-4~\cite{GPT4} to assess the GSB (Good/Same/Bad)~\cite{GSB} metric. This metric measures pairwise quality differences in the chain-of-thought (CoT) outputs generated before and after each training stage. Specifically, we randomly sample 200 user–item pairs from each dataset and generate recommendation reasons using three versions of the model: the original Qwen2.5 API (before any training), the adapted LLM (after recommendation post-training), and the refined LLM (after recommender-guided RL refinement). We then prompt GPT-4 to judge whether the reason quality improves (Good), remains comparable (Same), or degrades (Bad) after each stage. To mitigate positional bias, we additionally permute the input order and average the results. The evaluation outcomes are reported in Tab.~\ref{tab:GSB}.

Several observations can be drawn from the results. First, the recommendation adaptation post-training substantially enhances comprehensibility: on both MovieLens and Video Games, over 59\% of generated reasons are judged as improved compared to the vanilla Qwen2.5 API, while only around 20\% are considered degraded. This highlights the importance of aligning the LLM's reasoning process with recommendation objectives—without such adaptation, the LLM tends to produce generic descriptions that lack personalization and collaborative awareness. Besides, the subsequent refinement stage yields further improvements, with approximately 46--49.5\% of reasons rated as better than the already-adapted outputs. This demonstrates that the recommender-guided RL refinement effectively corrects hallucinated or overly generic reasons, producing more accurate and user-specific explanations. Together, these two stages progressively equip the LLM with stronger recommendation-oriented reasoning capabilities, enabling \method to produce comprehensible features for downstream recommendation tasks.

\subsubsection{Case Study}

\begin{figure}[t]
    \centering
    \subfigure[Generating reasonable features with tuned LLM.]{
        \includegraphics[width=\linewidth]{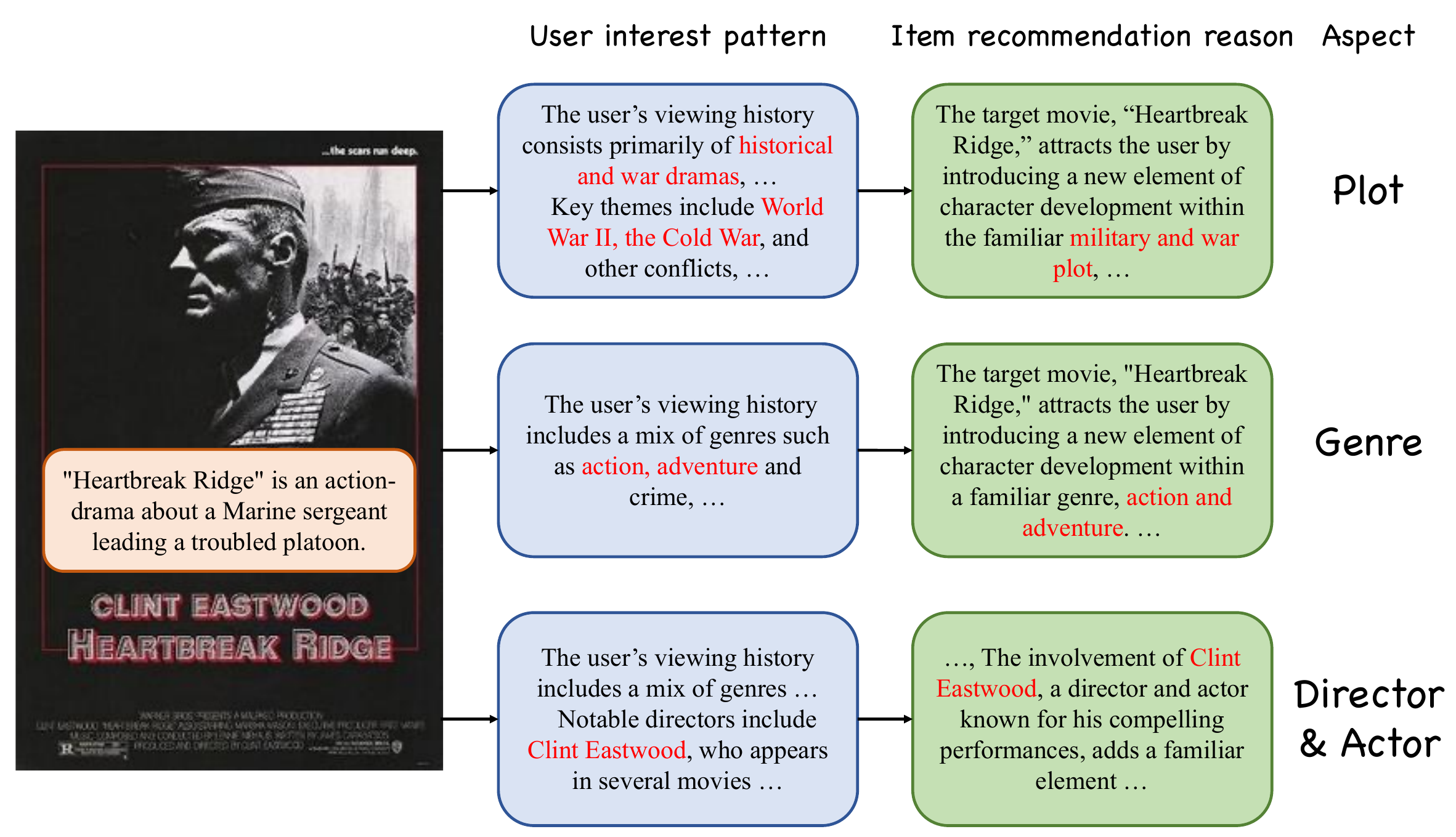}
        \label{fig:reason}
    }
    \subfigure[Matching pattern and reason with attention mechanism.]{
        \includegraphics[width=\linewidth]{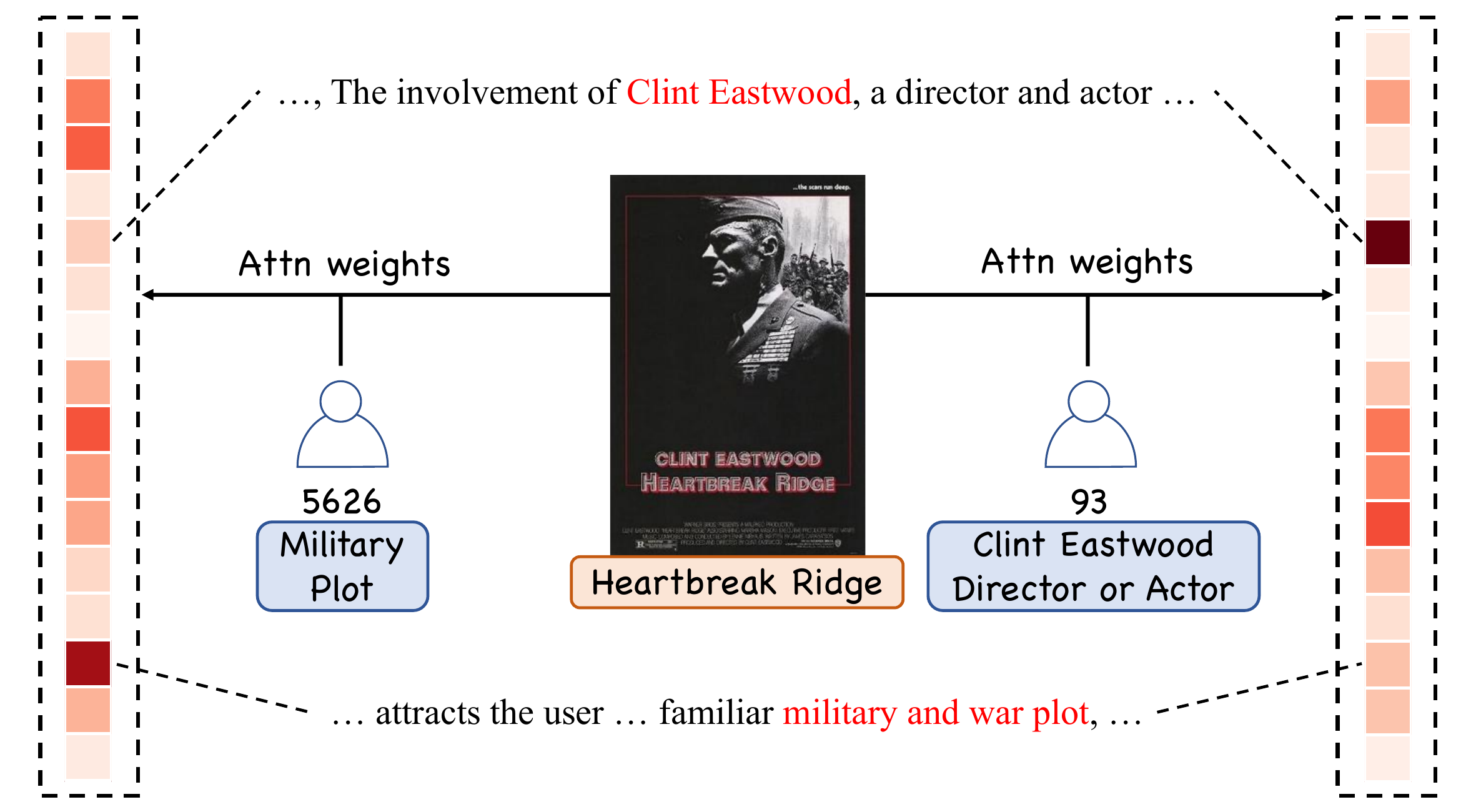}
        \label{fig:weight}
    }
    \caption{LLM-based \method enhances recommendation comprehensibility by introducing user patterns and item recommendation reasons as collaborative perspective features, and utilizing an attention mechanism to match the features.}
    \label{fig:case}
\end{figure}

To provide an intuitive illustration of how \method enhances recommendation comprehensibility, we conduct a qualitative case study. We select the same three users and the target movie (\textit{Heartbreak Ridge}) from the MovieLens dataset as presented in the motivating example (Fig.~\ref{fig:reasons}), and perform reasoning using the LLM trained within \method based on each user's distinct historical interaction sequence. The results are shown in Fig.~\ref{fig:case}.

As illustrated in Fig.~\ref{fig:reason}, \method successfully infers distinct user interest patterns from different historical behavior sequences, accurately capturing the key dimensions of each user's preferences (\textit{e.g.}, military-themed plots, action genres, and specific directors or actors). Crucially, the generated recommendation reasons are not mere item descriptions; instead, they are derived through in-depth reasoning over the intersection between the user's interest pattern and the item's content, reflecting a truly collaborative perspective. 

Furthermore, as shown in Fig.~\ref{fig:weight}, the attention mechanism effectively matches user interest patterns with the corresponding recommendation reasons in the item's reason list. The attention weights reveal which specific reasons are most relevant to each user's preferences, providing an explicit and interpretable alignment between user interests and item characteristics. This transparency allows end-users and system designers to understand \textit{why} a particular item is recommended, going beyond black-box scoring.

In summary, through the generation of personalized recommendation reason features and the attention-based matching mechanism between user interests and item reasons, \method significantly enhances the comprehensibility and transparency of the recommendation process, while contributing to improved prediction accuracy.

\section{Conclusion}

We propose \method, a novel framework that incorporates fine-tuned LLMs to generate collaborative-perspective content features and integrates them into recommendation, to enhance both performance and comprehensibility of recommendation.
\method first fine-tunes the LLM through adaptation for recommender systems. By training and introducing a recommender model, \method effectively evaluates the quality of the generated reasons and applies them in a recommender-guided refinement process to refine more accurate and personalized reasons. These features can be directly utilized in downstream recommendation tasks.
Extensive experiments demonstrate the superiority of \method over existing baselines, highlighting its ability to bridge the semantic-collaborative gap for content understanding comprehensive recommendation.

\begin{acks}
This work is partially sponsored by NSFC U24A20235 and "The Fundamental Research Funds for the Central Universities, Peking University".
\end{acks}

\newpage
\section*{GenAI Usage Disclosure}
During this study, we used Qwen2.5 as a core component of our model, as described in the Methods section. Generative AI was not used to assist with data processing or code generation. However, we used GPT-4 to address several questions encountered in the experiments. In the comprehensibility analysis, we additionally employed an LLM (GPT-4) as the judge for the GSB evaluation.

\bibliographystyle{ACM-Reference-Format}
\bibliography{reference}

@ARTICLE{CF,
  author={Koren, Yehuda and Bell, Robert and Volinsky, Chris},
  journal={Computer}, 
  title={Matrix Factorization Techniques for Recommender Systems}, 
  year={2009},
  volume={42},
  number={8},
  pages={30-37},
  doi={10.1109/MC.2009.263}}

@inproceedings{NCF,
author = {He, Xiangnan and Liao, Lizi and Zhang, Hanwang and Nie, Liqiang and Hu, Xia and Chua, Tat-Seng},
title = {Neural Collaborative Filtering},
year = {2017},
isbn = {9781450349130},
publisher = {International World Wide Web Conferences Steering Committee},
address = {Republic and Canton of Geneva, CHE},
url = {https://doi.org/10.1145/3038912.3052569},
doi = {10.1145/3038912.3052569},
booktitle = {Proceedings of the 26th International Conference on World Wide Web},
pages = {173–182},
numpages = {10},
location = {Perth, Australia},
series = {WWW '17}
}

@inproceedings{NGCF, series={SIGIR ’19},
   title={Neural Graph Collaborative Filtering},
   url={http://dx.doi.org/10.1145/3331184.3331267},
   DOI={10.1145/3331184.3331267},
   booktitle={Proceedings of the 42nd International ACM SIGIR Conference on Research and Development in Information Retrieval},
   publisher={ACM},
   author={Wang, Xiang and He, Xiangnan and Wang, Meng and Feng, Fuli and Chua, Tat-Seng},
   year={2019},
   month=jul, pages={165–174},
   collection={SIGIR ’19} }

@inproceedings{LightGCN,
author = {He, Xiangnan and Deng, Kuan and Wang, Xiang and Li, Yan and Zhang, YongDong and Wang, Meng},
title = {LightGCN: Simplifying and Powering Graph Convolution Network for Recommendation},
year = {2020},
isbn = {9781450380164},
publisher = {Association for Computing Machinery},
address = {New York, NY, USA},
url = {https://doi.org/10.1145/3397271.3401063},
doi = {10.1145/3397271.3401063},
booktitle = {Proceedings of the 43rd International ACM SIGIR Conference on Research and Development in Information Retrieval},
pages = {639–648},
numpages = {10},
location = {Virtual Event, China},
series = {SIGIR '20}
}

@inproceedings{HCCF,
author = {Xia, Lianghao and Huang, Chao and Xu, Yong and Zhao, Jiashu and Yin, Dawei and Huang, Jimmy},
title = {Hypergraph Contrastive Collaborative Filtering},
year = {2022},
isbn = {9781450387323},
publisher = {Association for Computing Machinery},
address = {New York, NY, USA},
url = {https://doi.org/10.1145/3477495.3532058},
doi = {10.1145/3477495.3532058},
booktitle = {Proceedings of the 45th International ACM SIGIR Conference on Research and Development in Information Retrieval},
pages = {70–79},
numpages = {10},
location = {Madrid, Spain},
series = {SIGIR '22}
}

@inproceedings{LLMRec,
author = {Kim, Sein and Kang, Hongseok and Choi, Seungyoon and Kim, Donghyun and Yang, Minchul and Park, Chanyoung},
title = {Large Language Models meet Collaborative Filtering: An Efficient All-round LLM-based Recommender System},
year = {2024},
isbn = {9798400704901},
publisher = {Association for Computing Machinery},
address = {New York, NY, USA},
url = {https://doi.org/10.1145/3637528.3671931},
doi = {10.1145/3637528.3671931},
booktitle = {Proceedings of the 30th ACM SIGKDD Conference on Knowledge Discovery and Data Mining},
pages = {1395–1406},
numpages = {12},
location = {Barcelona, Spain},
series = {KDD '24}
}

@inproceedings{M3CSR,
  title={A Multi-modal Modeling Framework for Cold-start Short-video Recommendation},
  author={Chen, Gaode and Sun, Ruina and Jiang, Yuezihan and Cao, Jiangxia and Zhang, Qi and Lin, Jingjian and Li, Han and Gai, Kun and Zhang, Xinghua},
  booktitle={Proceedings of the 18th ACM Conference on Recommender Systems},
  pages={391--400},
  year={2024}
}

@misc{SASRec,
      title={Self-Attentive Sequential Recommendation}, 
      author={Wang-Cheng Kang and Julian McAuley},
      year={2018},
      eprint={1808.09781},
      archivePrefix={arXiv},
      primaryClass={cs.IR},
      url={https://arxiv.org/abs/1808.09781}, 
}

@misc{NDCG,
      title={A Theoretical Analysis of NDCG Type Ranking Measures}, 
      author={Yining Wang and Liwei Wang and Yuanzhi Li and Di He and Tie-Yan Liu and Wei Chen},
      year={2013},
      eprint={1304.6480},
      archivePrefix={arXiv},
      primaryClass={cs.LG}
}

@misc{Adam,
  doi = {10.48550/ARXIV.1412.6980},
  url = {https://arxiv.org/abs/1412.6980},
  author = {Kingma, Diederik P. and Ba, Jimmy},
  title = {Adam: A Method for Stochastic Optimization},
  publisher = {arXiv},
  year = {2014},
  copyright = {arXiv.org perpetual, non-exclusive license}
}

@inproceedings{attention,
author = {Vaswani, Ashish and Shazeer, Noam and Parmar, Niki and Uszkoreit, Jakob and Jones, Llion and Gomez, Aidan N. and Kaiser, \L{}ukasz and Polosukhin, Illia},
title = {Attention is all you need},
year = {2017},
isbn = {9781510860964},
publisher = {Curran Associates Inc.},
address = {Red Hook, NY, USA},
booktitle = {Proceedings of the 31st International Conference on Neural Information Processing Systems},
pages = {6000–6010},
numpages = {11},
location = {Long Beach, California, USA},
series = {NIPS'17}
}

@INPROCEEDINGS{LC-REC,
  author={Zheng, Bowen and Hou, Yupeng and Lu, Hongyu and Chen, Yu and Zhao, Wayne Xin and Chen, Ming and Wen, Ji-Rong},
  booktitle={2024 IEEE 40th International Conference on Data Engineering (ICDE)}, 
  title={Adapting Large Language Models by Integrating Collaborative Semantics for Recommendation}, 
  year={2024},
  volume={},
  number={},
  pages={1435-1448},
  doi={10.1109/ICDE60146.2024.00118}}

@article{MovieLens,
author = {Harper, F. Maxwell and Konstan, Joseph A.},
title = {The MovieLens Datasets: History and Context},
year = {2015},
issue_date = {January 2016},
publisher = {Association for Computing Machinery},
address = {New York, NY, USA},
volume = {5},
number = {4},
issn = {2160-6455},
url = {https://doi.org/10.1145/2827872},
doi = {10.1145/2827872},
journal = {ACM Trans. Interact. Intell. Syst.},
month = {dec},
articleno = {19},
numpages = {19}
}

@article{Amazon,
  title={Bridging Language and Items for Retrieval and Recommendation},
  author={Hou, Yupeng and Li, Jiacheng and He, Zhankui and Yan, An and Chen, Xiusi and McAuley, Julian},
  journal={arXiv preprint arXiv:2403.03952},
  year={2024}
}

@inproceedings{Bert4Rec,
author = {Sun, Fei and Liu, Jun and Wu, Jian and Pei, Changhua and Lin, Xiao and Ou, Wenwu and Jiang, Peng},
title = {BERT4Rec: Sequential Recommendation with Bidirectional Encoder Representations from Transformer},
year = {2019},
isbn = {9781450369763},
publisher = {Association for Computing Machinery},
address = {New York, NY, USA},
url = {https://doi.org/10.1145/3357384.3357895},
doi = {10.1145/3357384.3357895},
booktitle = {Proceedings of the 28th ACM International Conference on Information and Knowledge Management},
pages = {1441–1450},
numpages = {10},
location = {Beijing, China},
series = {CIKM '19}
}

@inproceedings{SimRec,
author = {Xia, Lianghao and Huang, Chao and Shi, Jiao and Xu, Yong},
title = {Graph-less Collaborative Filtering},
year = {2023},
isbn = {9781450394161},
publisher = {Association for Computing Machinery},
address = {New York, NY, USA},
url = {https://doi.org/10.1145/3543507.3583196},
doi = {10.1145/3543507.3583196},
booktitle = {Proceedings of the ACM Web Conference 2023},
pages = {17–27},
numpages = {11},
location = {Austin, TX, USA},
series = {WWW '23}
}

@inproceedings{DuoRec,
author = {Qiu, Ruihong and Huang, Zi and Yin, Hongzhi and Wang, Zijian},
title = {Contrastive Learning for Representation Degeneration Problem in Sequential Recommendation},
year = {2022},
isbn = {9781450391320},
publisher = {Association for Computing Machinery},
address = {New York, NY, USA},
url = {https://doi.org/10.1145/3488560.3498433},
doi = {10.1145/3488560.3498433},
booktitle = {Proceedings of the Fifteenth ACM International Conference on Web Search and Data Mining},
pages = {813–823},
numpages = {11},
location = {Virtual Event, AZ, USA},
series = {WSDM '22}
}

@inproceedings{ICRec,
author = {Chen, Yongjun and Liu, Zhiwei and Li, Jia and McAuley, Julian and Xiong, Caiming},
title = {Intent Contrastive Learning for Sequential Recommendation},
year = {2022},
isbn = {9781450390965},
publisher = {Association for Computing Machinery},
address = {New York, NY, USA},
url = {https://doi.org/10.1145/3485447.3512090},
doi = {10.1145/3485447.3512090},
booktitle = {Proceedings of the ACM Web Conference 2022},
pages = {2172–2182},
numpages = {11},
location = {Virtual Event, Lyon, France},
series = {WWW '22}
}

@inproceedings{MAERec,
author = {Ye, Yaowen and Xia, Lianghao and Huang, Chao},
title = {Graph Masked Autoencoder for Sequential Recommendation},
year = {2023},
isbn = {9781450394086},
publisher = {Association for Computing Machinery},
address = {New York, NY, USA},
url = {https://doi.org/10.1145/3539618.3591692},
doi = {10.1145/3539618.3591692},
booktitle = {Proceedings of the 46th International ACM SIGIR Conference on Research and Development in Information Retrieval},
pages = {321–330},
numpages = {10},
location = {Taipei, Taiwan},
series = {SIGIR '23}
}

@inproceedings{StarGCN,
author = {Zhang, Jiani and Shi, Xingjian and Zhao, Shenglin and King, Irwin},
title = {STAR-GCN: stacked and reconstructed graph convolutional networks for recommender systems},
year = {2019},
isbn = {9780999241141},
publisher = {AAAI Press},
booktitle = {Proceedings of the 28th International Joint Conference on Artificial Intelligence},
pages = {4264–4270},
numpages = {7},
location = {Macao, China},
series = {IJCAI'19}
}

@inproceedings{GMCF,
author = {Su, Yixin and Zhang, Rui and M. Erfani, Sarah and Gan, Junhao},
title = {Neural Graph Matching based Collaborative Filtering},
year = {2021},
isbn = {9781450380379},
publisher = {Association for Computing Machinery},
address = {New York, NY, USA},
url = {https://doi.org/10.1145/3404835.3462833},
doi = {10.1145/3404835.3462833},
booktitle = {Proceedings of the 44th International ACM SIGIR Conference on Research and Development in Information Retrieval},
pages = {849–858},
numpages = {10},
location = {Virtual Event, Canada},
series = {SIGIR '21}
}

@inproceedings{CASER,
author = {Tang, Jiaxi and Wang, Ke},
title = {Personalized Top-N Sequential Recommendation via Convolutional Sequence Embedding},
year = {2018},
isbn = {9781450355810},
publisher = {Association for Computing Machinery},
address = {New York, NY, USA},
url = {https://doi.org/10.1145/3159652.3159656},
doi = {10.1145/3159652.3159656},
booktitle = {Proceedings of the Eleventh ACM International Conference on Web Search and Data Mining},
pages = {565–573},
numpages = {9},
location = {Marina Del Rey, CA, USA},
series = {WSDM '18}
}

@article{SRGNN, title={Session-Based Recommendation with Graph Neural Networks}, volume={33}, url={https://ojs.aaai.org/index.php/AAAI/article/view/3804}, DOI={10.1609/aaai.v33i01.3301346}, abstractNote={&lt;p&gt;The problem of session-based recommendation aims to predict user actions based on anonymous sessions. Previous methods model a session as a sequence and estimate user representations besides item representations to make recommendations. Though achieved promising results, they are insufficient to obtain accurate user vectors in sessions and neglect complex transitions of items. To obtain accurate item embedding and take complex transitions of items into account, we propose a novel method, i.e. &lt;em&gt;Session-based Recommendation with Graph Neural Networks&lt;/em&gt;, SR-GNN for brevity. In the proposed method, session sequences are modeled as graphstructured data. Based on the session graph, GNN can capture complex transitions of items, which are difficult to be revealed by previous conventional sequential methods. Each session is then represented as the composition of the global preference and the current interest of that session using an attention network. Extensive experiments conducted on two real datasets show that SR-GNN evidently outperforms the state-of-the-art session-based recommendation methods consistently.&lt;/p&gt;}, number={01}, journal={Proceedings of the AAAI Conference on Artificial Intelligence}, author={Wu, Shu and Tang, Yuyuan and Zhu, Yanqiao and Wang, Liang and Xie, Xing and Tan, Tieniu}, year={2019}, month={Jul.}, pages={346-353} }

@inproceedings{MAKE,
author = {Sheng, Xiang-Rong and Yang, Feifan and Gong, Litong and Wang, Biao and Chan, Zhangming and Zhang, Yujing and Cheng, Yueyao and Zhu, Yong-Nan and Ge, Tiezheng and Zhu, Han and Jiang, Yuning and Xu, Jian and Zheng, Bo},
title = {Enhancing Taobao Display Advertising with Multimodal Representations: Challenges, Approaches and Insights},
year = {2024},
isbn = {9798400704369},
publisher = {Association for Computing Machinery},
address = {New York, NY, USA},
url = {https://doi.org/10.1145/3627673.3680068},
doi = {10.1145/3627673.3680068},
booktitle = {Proceedings of the 33rd ACM International Conference on Information and Knowledge Management},
pages = {4858–4865},
numpages = {8},
location = {Boise, ID, USA},
series = {CIKM '24}
}

@inproceedings{BERT,
    title = "{BERT}: Pre-training of Deep Bidirectional Transformers for Language Understanding",
    author = "Devlin, Jacob  and
      Chang, Ming-Wei  and
      Lee, Kenton  and
      Toutanova, Kristina",
    editor = "Burstein, Jill  and
      Doran, Christy  and
      Solorio, Thamar",
    booktitle = "Proceedings of the 2019 Conference of the North {A}merican Chapter of the Association for Computational Linguistics: Human Language Technologies, Volume 1 (Long and Short Papers)",
    month = jun,
    year = "2019",
    address = "Minneapolis, Minnesota",
    publisher = "Association for Computational Linguistics",
    url = "https://aclanthology.org/N19-1423/",
    doi = "10.18653/v1/N19-1423",
    pages = "4171--4186"
}

@misc{ZESRec,
      title={Zero-Shot Recommender Systems}, 
      author={Hao Ding and Yifei Ma and Anoop Deoras and Yuyang Wang and Hao Wang},
      year={2021},
      eprint={2105.08318},
      archivePrefix={arXiv},
      primaryClass={cs.LG},
      url={https://arxiv.org/abs/2105.08318}, 
}

@inproceedings{UniSRec,
author = {Hou, Yupeng and Mu, Shanlei and Zhao, Wayne Xin and Li, Yaliang and Ding, Bolin and Wen, Ji-Rong},
title = {Towards Universal Sequence Representation Learning for Recommender Systems},
year = {2022},
isbn = {9781450393850},
publisher = {Association for Computing Machinery},
address = {New York, NY, USA},
url = {https://doi.org/10.1145/3534678.3539381},
doi = {10.1145/3534678.3539381},
booktitle = {Proceedings of the 28th ACM SIGKDD Conference on Knowledge Discovery and Data Mining},
pages = {585–593},
numpages = {9},
location = {Washington DC, USA},
series = {KDD '22}
}

@inproceedings{RecFORMER,
author = {Li, Jiacheng and Wang, Ming and Li, Jin and Fu, Jinmiao and Shen, Xin and Shang, Jingbo and McAuley, Julian},
title = {Text Is All You Need: Learning Language Representations for Sequential Recommendation},
year = {2023},
isbn = {9798400701030},
publisher = {Association for Computing Machinery},
address = {New York, NY, USA},
url = {https://doi.org/10.1145/3580305.3599519},
doi = {10.1145/3580305.3599519},
booktitle = {Proceedings of the 29th ACM SIGKDD Conference on Knowledge Discovery and Data Mining},
pages = {1258–1267},
numpages = {10},
location = {Long Beach, CA, USA},
series = {KDD '23}
}

@article{VBPR,
	title = {{VBPR}: {Visual} {Bayesian} {Personalized} {Ranking} from {Implicit} {Feedback}},
	volume = {30},
	url = {https://ojs.aaai.org/index.php/AAAI/article/view/9973},
	doi = {10.1609/aaai.v30i1.9973},
	number = {1},
	journal = {Proceedings of the AAAI Conference on Artificial Intelligence},
	author = {He, Ruining and McAuley, Julian},
	month = feb,
	year = {2016},
}

@inproceedings{SeRALM,
author = {Ren, Yankun and Chen, Zhongde and Yang, Xinxing and Li, Longfei and Jiang, Cong and Cheng, Lei and Zhang, Bo and Mo, Linjian and Zhou, Jun},
title = {Enhancing Sequential Recommenders with Augmented Knowledge from Aligned Large Language Models},
year = {2024},
isbn = {9798400704314},
publisher = {Association for Computing Machinery},
address = {New York, NY, USA},
url = {https://doi.org/10.1145/3626772.3657782},
doi = {10.1145/3626772.3657782},
booktitle = {Proceedings of the 47th International ACM SIGIR Conference on Research and Development in Information Retrieval},
pages = {345–354},
numpages = {10},
location = {Washington DC, USA},
series = {SIGIR '24}
}

@inproceedings{TALLRec,
author = {Bao, Keqin and Zhang, Jizhi and Zhang, Yang and Wang, Wenjie and Feng, Fuli and He, Xiangnan},
title = {TALLRec: An Effective and Efficient Tuning Framework to Align Large Language Model with Recommendation},
year = {2023},
isbn = {9798400702419},
publisher = {Association for Computing Machinery},
address = {New York, NY, USA},
url = {https://doi.org/10.1145/3604915.3608857},
doi = {10.1145/3604915.3608857},
booktitle = {Proceedings of the 17th ACM Conference on Recommender Systems},
pages = {1007–1014},
numpages = {8},
location = {Singapore, Singapore},
series = {RecSys '23}
}

@misc{LlamaRec,
      title={LlamaRec: Two-Stage Recommendation using Large Language Models for Ranking}, 
      author={Zhenrui Yue and Sara Rabhi and Gabriel de Souza Pereira Moreira and Dong Wang and Even Oldridge},
      year={2023},
      eprint={2311.02089},
      archivePrefix={arXiv},
      primaryClass={cs.IR},
      url={https://arxiv.org/abs/2311.02089}, 
}

@inproceedings{BinLLM,
    title = "Text-like Encoding of Collaborative Information in Large Language Models for Recommendation",
    author = "Zhang, Yang  and
      Bao, Keqin  and
      Yan, Ming  and
      Wang, Wenjie  and
      Feng, Fuli  and
      He, Xiangnan",
    editor = "Ku, Lun-Wei  and
      Martins, Andre  and
      Srikumar, Vivek",
    booktitle = "Proceedings of the 62nd Annual Meeting of the Association for Computational Linguistics (Volume 1: Long Papers)",
    month = aug,
    year = "2024",
    address = "Bangkok, Thailand",
    publisher = "Association for Computational Linguistics",
    url = "https://aclanthology.org/2024.acl-long.497/",
    doi = "10.18653/v1/2024.acl-long.497",
    pages = "9181--9191"
}

@ARTICLE{CoLLM,
  author={Zhang, Yang and Feng, Fuli and Zhang, Jizhi and Bao, Keqin and Wang, Qifan and He, Xiangnan},
  journal={IEEE Transactions on Knowledge and Data Engineering}, 
  title={CoLLM: Integrating Collaborative Embeddings Into Large Language Models for Recommendation}, 
  year={2025},
  volume={37},
  number={5},
  pages={2329-2340},
  doi={10.1109/TKDE.2025.3540912}}

@inproceedings{KAR,
author = {Xi, Yunjia and Liu, Weiwen and Lin, Jianghao and Cai, Xiaoling and Zhu, Hong and Zhu, Jieming and Chen, Bo and Tang, Ruiming and Zhang, Weinan and Yu, Yong},
title = {Towards Open-World Recommendation with Knowledge Augmentation from Large Language Models},
year = {2024},
isbn = {9798400705052},
publisher = {Association for Computing Machinery},
address = {New York, NY, USA},
url = {https://doi.org/10.1145/3640457.3688104},
doi = {10.1145/3640457.3688104},
booktitle = {Proceedings of the 18th ACM Conference on Recommender Systems},
pages = {12–22},
numpages = {11},
location = {Bari, Italy},
series = {RecSys '24}
}

@inproceedings{HiT,
author = {Xia, Yu and Zhong, Rui and Gu, Hao and Yang, Wei and Lu, Chi and Jiang, Peng and Gai, Kun},
title = {Hierarchical Tree Search-based User Lifelong Behavior Modeling on Large Language Model},
year = {2025},
isbn = {9798400715921},
publisher = {Association for Computing Machinery},
address = {New York, NY, USA},
url = {https://doi.org/10.1145/3726302.3729995},
doi = {10.1145/3726302.3729995},
booktitle = {Proceedings of the 48th International ACM SIGIR Conference on Research and Development in Information Retrieval},
pages = {1758–1767},
numpages = {10},
location = {Padua, Italy},
series = {SIGIR '25}
}

@misc{LEARN,
	title = {{LEARN}: {Knowledge} {Adaptation} from {Large} {Language} {Model} to {Recommendation} for {Practical} {Industrial} {Application}},
	shorttitle = {{LEARN}},
	url = {http://arxiv.org/abs/2405.03988},
	doi = {10.48550/arXiv.2405.03988},
	language = {en},
	urldate = {2025-07-11},
	publisher = {arXiv},
	author = {Jia, Jian and Wang, Yipei and Li, Yan and Chen, Honggang and Bai, Xuehan and Liu, Zhaocheng and Liang, Jian and Chen, Quan and Li, Han and Jiang, Peng and Gai, Kun},
	month = dec,
	year = {2024},
	note = {arXiv:2405.03988 [cs]},
}

@misc{ReasoningRec,
      title={ReasoningRec: Bridging Personalized Recommendations and Human-Interpretable Explanations through LLM Reasoning}, 
      author={Millennium Bismay and Xiangjue Dong and James Caverlee},
      year={2024},
      eprint={2410.23180},
      archivePrefix={arXiv},
      primaryClass={cs.IR},
      url={https://arxiv.org/abs/2410.23180}, 
}

@inproceedings{PAM,
author = {Luo, Yunze and Jiang, Yuezihan and Jiang, Yinjie and Chen, Gaode and Wang, Jingchi and Bian, Kaigui and Li, Peiyi and Zhang, Qi},
title = {Online Item Cold-Start Recommendation with Popularity-Aware Meta-Learning},
year = {2025},
isbn = {9798400712456},
publisher = {Association for Computing Machinery},
address = {New York, NY, USA},
url = {https://doi.org/10.1145/3690624.3709336},
doi = {10.1145/3690624.3709336},
booktitle = {Proceedings of the 31st ACM SIGKDD Conference on Knowledge Discovery and Data Mining V.1},
pages = {927–937},
numpages = {11},
location = {Toronto ON, Canada},
series = {KDD '25}
}

@inproceedings{WaD,
author = {Cheng, Heng-Tze and Koc, Levent and Harmsen, Jeremiah and Shaked, Tal and Chandra, Tushar and Aradhye, Hrishi and Anderson, Glen and Corrado, Greg and Chai, Wei and Ispir, Mustafa and Anil, Rohan and Haque, Zakaria and Hong, Lichan and Jain, Vihan and Liu, Xiaobing and Shah, Hemal},
title = {Wide \& Deep Learning for Recommender Systems},
year = {2016},
isbn = {9781450347952},
publisher = {Association for Computing Machinery},
address = {New York, NY, USA},
url = {https://doi.org/10.1145/2988450.2988454},
doi = {10.1145/2988450.2988454},
booktitle = {Proceedings of the 1st Workshop on Deep Learning for Recommender Systems},
pages = {7–10},
numpages = {4},
location = {Boston, MA, USA},
series = {DLRS 2016}
}

@inproceedings{S3Rec,
author = {Zhou, Kun and Wang, Hui and Zhao, Wayne Xin and Zhu, Yutao and Wang, Sirui and Zhang, Fuzheng and Wang, Zhongyuan and Wen, Ji-Rong},
title = {S3-Rec: Self-Supervised Learning for Sequential Recommendation with Mutual Information Maximization},
year = {2020},
isbn = {9781450368599},
publisher = {Association for Computing Machinery},
address = {New York, NY, USA},
url = {https://doi.org/10.1145/3340531.3411954},
doi = {10.1145/3340531.3411954},
booktitle = {Proceedings of the 29th ACM International Conference on Information \& Knowledge Management},
pages = {1893–1902},
numpages = {10},
location = {Virtual Event, Ireland},
series = {CIKM '20}
}

@article{Hallucination,
author = {Ji, Ziwei and Lee, Nayeon and Frieske, Rita and Yu, Tiezheng and Su, Dan and Xu, Yan and Ishii, Etsuko and Bang, Ye Jin and Madotto, Andrea and Fung, Pascale},
title = {Survey of Hallucination in Natural Language Generation},
year = {2023},
issue_date = {December 2023},
publisher = {Association for Computing Machinery},
address = {New York, NY, USA},
volume = {55},
number = {12},
issn = {0360-0300},
url = {https://doi.org/10.1145/3571730},
doi = {10.1145/3571730},
journal = {ACM Comput. Surv.},
month = mar,
articleno = {248},
numpages = {38}
}

@misc{GRPO,
      title={DeepSeekMath: Pushing the Limits of Mathematical Reasoning in Open Language Models}, 
      author={Zhihong Shao and Peiyi Wang and Qihao Zhu and Runxin Xu and Junxiao Song and Xiao Bi and Haowei Zhang and Mingchuan Zhang and Y. K. Li and Y. Wu and Daya Guo},
      year={2024},
      eprint={2402.03300},
      archivePrefix={arXiv},
      primaryClass={cs.CL},
      url={https://arxiv.org/abs/2402.03300}, 
}

@inproceedings{DIN,
author = {Zhou, Guorui and Zhu, Xiaoqiang and Song, Chenru and Fan, Ying and Zhu, Han and Ma, Xiao and Yan, Yanghui and Jin, Junqi and Li, Han and Gai, Kun},
title = {Deep Interest Network for Click-Through Rate Prediction},
year = {2018},
isbn = {9781450355520},
publisher = {Association for Computing Machinery},
address = {New York, NY, USA},
url = {https://doi.org/10.1145/3219819.3219823},
doi = {10.1145/3219819.3219823},
booktitle = {Proceedings of the 24th ACM SIGKDD International Conference on Knowledge Discovery \& Data Mining},
pages = {1059–1068},
numpages = {10},
location = {London, United Kingdom},
series = {KDD '18}
}

@misc{Qwen,
      title={Qwen2.5 Technical Report}, 
      author={Qwen et al.},
      year={2025},
      eprint={2412.15115},
      archivePrefix={arXiv},
      primaryClass={cs.CL},
      url={https://arxiv.org/abs/2412.15115}, 
}

@misc{AdamW,
      title={Decoupled Weight Decay Regularization}, 
      author={Ilya Loshchilov and Frank Hutter},
      year={2019},
      eprint={1711.05101},
      archivePrefix={arXiv},
      primaryClass={cs.LG},
      url={https://arxiv.org/abs/1711.05101}, 
}

@misc{LONGER,
      title={LONGER: Scaling Up Long Sequence Modeling in Industrial Recommenders}, 
      author={Zheng Chai and Qin Ren and Xijun Xiao and Huizhi Yang and Bo Han and Sijun Zhang and Di Chen and Hui Lu and Wenlin Zhao and Lele Yu and Xionghang Xie and Shiru Ren and Xiang Sun and Yaocheng Tan and Peng Xu and Yuchao Zheng and Di Wu},
      year={2025},
      eprint={2505.04421},
      archivePrefix={arXiv},
      primaryClass={cs.IR},
      url={https://arxiv.org/abs/2505.04421}, 
}

@inproceedings{social,
author = {Zhou, Xiangmin and Chen, Lei and He, Chengkun and Wu, Junfeng and Zhou, Weiyi and Shao, Jie and Zhang, Yanchun},
title = {A Responsible and Extendable Context-Aware Recommender System},
year = {2025},
isbn = {9798400713316},
publisher = {Association for Computing Machinery},
address = {New York, NY, USA},
url = {https://doi.org/10.1145/3701716.3715164},
doi = {10.1145/3701716.3715164},
booktitle = {Companion Proceedings of the ACM on Web Conference 2025},
pages = {2955–2958},
numpages = {4},
location = {Sydney NSW, Australia},
series = {WWW '25}
}

@misc{reasoning,
      title={Do Larger Language Models Imply Better Generalization? A Pretraining Scaling Law for Implicit Reasoning}, 
      author={Xinyi Wang and Shawn Tan and Mingyu Jin and William Yang Wang and Rameswar Panda and Yikang Shen},
      year={2025},
      eprint={2504.03635},
      archivePrefix={arXiv},
      primaryClass={cs.AI},
      url={https://arxiv.org/abs/2504.03635}, 
}

@inproceedings{AlphaRec,
  author       = {Leheng Sheng and
                  An Zhang and
                  Yi Zhang and
                  Yuxin Chen and
                  Xiang Wang and
                  Tat{-}Seng Chua},
  title        = {Language Representations Can be What Recommenders Need: Findings and
                  Potentials},
  booktitle    = {ICLR},
  year         = {2025}
}

@misc{RLMRec,
      title={Language Representations Can be What Recommenders Need: Findings and Potentials}, 
      author={Leheng Sheng and An Zhang and Yi Zhang and Yuxin Chen and Xiang Wang and Tat-Seng Chua},
      year={2025},
      eprint={2407.05441},
      archivePrefix={arXiv},
      primaryClass={cs.IR},
      url={https://arxiv.org/abs/2407.05441}, 
}

@inproceedings{GSB,
    title = "Personalized Large Language Model Assistant with Evolving Conditional Memory",
    author = "Yuan, Ruifeng  and
      Sun, Shichao  and
      Li, Yongqi  and
      Wang, Zili  and
      Cao, Ziqiang  and
      Li, Wenjie",
    editor = "Rambow, Owen  and
      Wanner, Leo  and
      Apidianaki, Marianna  and
      Al-Khalifa, Hend  and
      Eugenio, Barbara Di  and
      Schockaert, Steven",
    booktitle = "Proceedings of the 31st International Conference on Computational Linguistics",
    month = jan,
    year = "2025",
    address = "Abu Dhabi, UAE",
    publisher = "Association for Computational Linguistics",
    url = "https://aclanthology.org/2025.coling-main.254/",
    pages = "3764--3777"
}

@misc{GPT4,
      title={GPT-4 Technical Report}, 
      author={OpenAI et al.},
      year={2024},
      eprint={2303.08774},
      archivePrefix={arXiv},
      primaryClass={cs.CL},
      url={https://arxiv.org/abs/2303.08774}, 
}

\end{document}